%% file: main.tex
\documentclass[letterpaper,twocolumn,10pt]{article}
\usepackage{usenix-2020-09}

\usepackage{tikz}
\usepackage{amsmath}
\usepackage{amssymb}
\usepackage{nomencl}
\usepackage{siunitx}
\usepackage{caption}
\usepackage{subcaption}
\usepackage{multirow}
\usepackage{hyperref}
\usepackage{cleveref}
\usepackage{bm}

\crefformat{section}{\S#2#1#3} 
\crefformat{subsection}{\S#2#1#3}
\crefformat{subsubsection}{\S#2#1#3}
\crefformat{paragraph}{\S#2#1#3}

\usepackage{enumitem}

\newcommand*\circled[1]{\tikz[baseline=(char.base)]{
            \node[shape=circle,draw,inner sep=2pt] (char) {#1};}}

\begin{document}

\date{}

\title{Taming and Controlling Performance and Energy Trade-offs Automatically in Network Applications}

\author{Han Dong, Yara Awad, Sanjay Arora*, Orran Krieger, Jonathan Appavoo\\
  \textit{Boston University, Red Hat*}
  }
\maketitle

\begin{abstract}
\input{abstract2}
\end{abstract}

\input{intro5}

\input{study2}

\input{model2}

\input{bayop}

\input{related}

\input{conclusion}
\bibliographystyle{plain}
\bibliography{references}

\clearpage
\appendix
\input{appendix}

\end{document}

%% file: abstract2.tex
In this paper, we demonstrate that a server running a single latency-sensitive application can be treated as a black box to reduce energy consumption while meeting an SLA target. We find that when the mean offered load is stable, one can find the ``sweet spot'' settings in packet batching (via interrupt coalescing) and controlling the processing rate (DVFS) that represents optimal trade-offs in the interactions of the software stack and hardware with the arrival rate and composition of requests currently being served. Trying a few combinations of settings on the live system, an example Bayesian optimizer can find settings that reduce the energy consumption to meet a desired tail latency for the current load.

This research demonstrates that: 1) without software changes, dramatic energy savings (up to 60\%) can be achieved across diverse hardware systems if one controls batching and processing rate, 2) specialized research OSes that have been developed for performance can achieve more than 2x better energy efficiency than general-purpose OSes, and 3) a controller, agnostic to the application and system, can easily find energy-efficient settings for the offered load that meets SLA objectives.

%% file: intro5.tex
\section{Introduction}
Today latency-sensitive cloud applications\footnote{Examples of latency-sensitive cloud applications include kev-value stores, search, and image and speech recognition.  The execution of such applications must often meet a specific performance target expressed as a Service Level Agreement (SLA). A common SLA is a 99\% tail latency requirement -- Eg. 99\% of all requests must be completed within some time limit.} 
play a critical role; in many cases, fleets of servers are dedicated to running a single instance of these applications~\cite{ixcp, heracles, PerAppPower, twine, twittermcd}. 
Researchers have shown that it is worth exploiting techniques to bypass the kernel and design highly specialized software stacks that combine a purpose-built library OS with these applications to improve their performance~\cite{ix, arrakis, zygos, shenango, rumpkernel, aliraza, unikernels, scalingmcdfacebook, arachne, mtcp, sandstorm, affinityaccept, flexnic, mica, seda}. 
As global data center energy use continues to rise~\cite{gupta2020chasing, NLP-energy,warehouse-power,nature1}, it is critical to find ways to meet the challenging requirements of these applications while reducing their energy use. 

Studies of latency-sensitive applications have shown that they experience stable mean demand curves.  
These curves show gradual changes in offered loads over extended periods, ranging from multiple hours to days. Such stability arises from recurring diurnal patterns and use of load balancers~\cite{scalingmcdfacebook, twittermcd, netflixmcd, redditmcd}.
Generally, these studies suggest that for a particular service there exists a stable mean arrival rate and composition of requests over some time scale.

This load stability (i.e. request rates and composition of requests) offers opportunities to meet SLA objectives while reducing energy use. Specifically, queuing theory suggests that the slack between request arrival, service time, and the SLA can be leveraged to improve energy efficiency. For example, induced queuing can amortize per-packet overhead to improve coalescing and processing efficiency~\cite{mootaz}, and even introduce idle periods in which the system can enter low-energy sleep states~\cite{slowdownorsleep}.

However, for a specific offered load, application, operating system (OS), and hardware, the most energy-efficient way to meet the SLA objective is specific to how the exact combination of software and hardware interacts.  
For example, queuing and processing rate settings that mimic a ``race-to-idle'' policy, executing as fast as possible to create the greatest amount of idle time to spend in a deep sleep state (that may flush CPU caches), maybe the right choice. 
It is, however, also possible that for the combination of hardware and software being used, it is better to choose a setting that mimics a "pace-to-idle" policy, executing more slowly and either entering a light sleep state or not entering a sleep state altogether~\cite{pacingtoidle}.  

Our research adds to the body of work on energy management\cite{slowdownorsleep,dreamweaver,dynsleep,pacingtoidle} by demonstrating that one can exploit stability in system behavior to efficiently find queuing and CPU processing rate settings to meet a tail latency target while reducing energy consumption.
We explore three basic conjectures:
\begin{enumerate}

\item There is a combination of queuing and CPU frequencies for a
  particular offered load and system (application, OS, hardware) that yields ``sweet spots'' where one can
  achieve an acceptable latency distribution while significantly reducing energy
  consumption.

\item Despite complex interactions between software and hardware, the ``sweet spot'' setting for
  a system and load are stable and, once found, will continue to yield good behavior \textit{if} queuing and CPU frequencies are fixed (i.e., not dynamically changed by the OS).

\item A system's response to changes in the queuing and CPU frequencies, at a fixed mean offered load, is well-behaved such that
  it is possible to use a generic black-box search strategy to quickly find a ``sweet spot'' setting on a running system.\footnote{Such an approach has the potential to be
  universal as it operates at runtime on the entire system and does not depend on tables of parameters, prior training, or profiling.}

\end{enumerate}

We explore the first two conjectures in an experimental study (\cref{sec:study}) on two applications across two distinct OSes: Linux and an OS specialized for latency-sensitive applications (EbbRT~\cite{ebbrt}). Similar to Co-PI~\cite{9248059}, we use existing hardware mechanisms: network interrupt coalescing (ITR-delay~\cite{intelinterruptmoderation}) and dynamic voltage frequency scaling~\cite{cpufreq_governor} (DVFS) to externally sweep queuing and CPU frequency on the server for a fixed set of offered load. Our study explored up to 340 combinations of ITR-delay, and DVFS and found "sweet spots" that both improved performance by 60\% while also lowering energy use by 50\% (\cref{sec:closed_energy}) in closed-loop applications. For open-loop applications (\cref{sec:open1}, \cref{sec:open2}), these mechanisms can lower energy use by 76\% in Linux (in contrast to its \textit{ondemand} DVFS policy)\footnote{We've studied the available Linux governors and found \textit{ondemand} to the most appropriate for these workloads.}  while meeting SLA objectives. 
We find that the specialized system has not only much better performance but also achieves a \textbf{further} 43\% reduction in energy. 
Most importantly, we found that, while the settings differ, the general purpose and specialized system have similar responses to changes, suggesting one could formally capture this common structure.

This common structure led us to the third conjecture -- it is plausible to use a generic search strategy to dynamically find energy-efficient ITR-delay, and DVFS settings for a given offered load. 
We successfully model (\cref{sec:model}) our experimental data to capture latency and energy profiles across both OSes. 
The accuracy of our model fit suggests that  a generic black-box-based controller can be used.
We then  built a prototype controller (\cref{sec:cachetrace}) using 
an established machine learning technique, Bayesian optimization~\cite{garnett_bayesoptbook_2022}, and we illustrate its use in exploiting the stable mean demand curve of a publicly available key-value trace~\cite{cacheWorkload-OSDI20} to save up to 60\% in server energy. 
Note that the goal of the prototype is to validate that a black box approach is possible; issues like how and when to trigger search are not studied. 
Finally (\cref{sec:tailbench}) we demonstrate the generality of our approach,  finding savings up to 36\% on applications different from our study (Tailbench~\cite{tailbench}) and on radically different hardware platforms released almost a decade apart 
(i.e. Intel E5-2640-released Q1'12 and Ampere ARMv8 released Q4'21) with different interrupt coalescing mechanisms. 

This work shows that in environments where: 1) dedicated servers are used for critical cloud applications and, 2) there is significant stability (on the order of minutes) in offered load: 
\begin{enumerate}
   \item  There are dramatic energy savings possible if one controls queuing and CPU frequency outside the OS for an offered demand; controls that can be applied to a general purpose OS like Linux with no changes.
   
   \item  Today's specialized research systems that have been developed for performance achieve dramatically better energy use than general purpose system when run baremetal. 
   
    \item A black-box-based controller can be built to exploit the stable demand curves of latency-sensitive applications to find energy-efficient "sweet spots" that are apply across a range of applications, operating systems and hardware. 
\end{enumerate}
The rest of our paper is structured as follows: \cref{sec:study} details our study and some key experimental findings, \cref{sec:model} presents a subset of our modeling results as motivation towards the design and evaluation of our controller in \cref{sec:cachetrace}. We then present related works in \cref{sec:related} and conclude in \cref{sec:conc}.

%% file: study2.tex
\section{Energy Study}
\label{sec:study}
We designed this study to validate our conjectures that externally manipulating queuing and CPU frequency can yield a diverse space for exploring energy-efficient "sweet spots". To our knowledge, this study is the first to conduct a study of interrupt coalescing, CPU frequency combinations across two distinct OSes running baremetal, and with a variety of network applications shown in~\cref{table:wrkcfgs}.

\renewcommand{\tabcolsep}{2pt}
\begin{table}[t]
\small
\begin{tabular}{|c|c|c|c|c|}
    \hline
  OS & App & Network Loads & Loop & Type\\
  \hline
     & NetPIPE & 64B, 8KB, 64KB, 512KB & Closed & OS\\ \cline{2-5}
     Linux, & NodeJS & N/A & Closed & App \\ \cline{2-5}
   EbbRT & Memcached & 200K, 400K, 600K QPS & Open & OS\\ \cline{2-5}
   & Silo & 50K, 100K, 200K QPS & Open & App\\ \hline
\end{tabular}
\vspace{-10pt}
\caption{
Operating system (OS), application, and network configurations.
\textbf{Network Loads} reflect mean values: requests-per-second \textit{(QPS)} or message sizes \textit{(KB)}. \textbf{Type} indicates whether an application is more reliant on application processing or OS processing.}
\label{table:wrkcfgs}
\end{table}

\subsection{Study Setup}
Our infrastructure consists of seven nodes, featuring a mix of 16-core Intel(R) Xeon(R) CPU E5-2690 @ 2.90GHz with 126 GB RAM and 12-core Intel(R) Xeon(R) CPU E5-2630L v2 @ 2.40GHz processors with 256 GB RAM, all equipped with Intel 82599ES 10-Gigabit SFI/SFP+ NICs. The single node used for booting into both baremetal EbbRT and Linux includes a 16-core Intel(R) Xeon(R) CPU E5-2690 @ 2.90GHz, 126 GB RAM, and an Intel 82599ES 10-Gigabit SFI/SFP+ NIC. Ensuring hardware parity between Linux and EbbRT, we carefully configured IA-32 Architectural MSRs, processor-specific MSRs (refer to Tables 35-2 and 35-18 in~\cite{intel_msr}), and NIC features, including disabled direct-cache injection (DCA), enabled receive-side scaling (RSS) for multi-core packet processing distribution, and enabled hardware checksum offloading. We matched NIC transmit and receive descriptors and write-back thresholds for packet transmissions. Additionally, to minimize system noise, hyperthreads and TurboBoost are disabled on all processors. We excluded TurboBoost due to reported energy anomalies when used with different sleep states~\cite{slowdownorsleep}. 
\begin{figure*}[hbt!]
\centering
\includegraphics[width=1.0\textwidth]{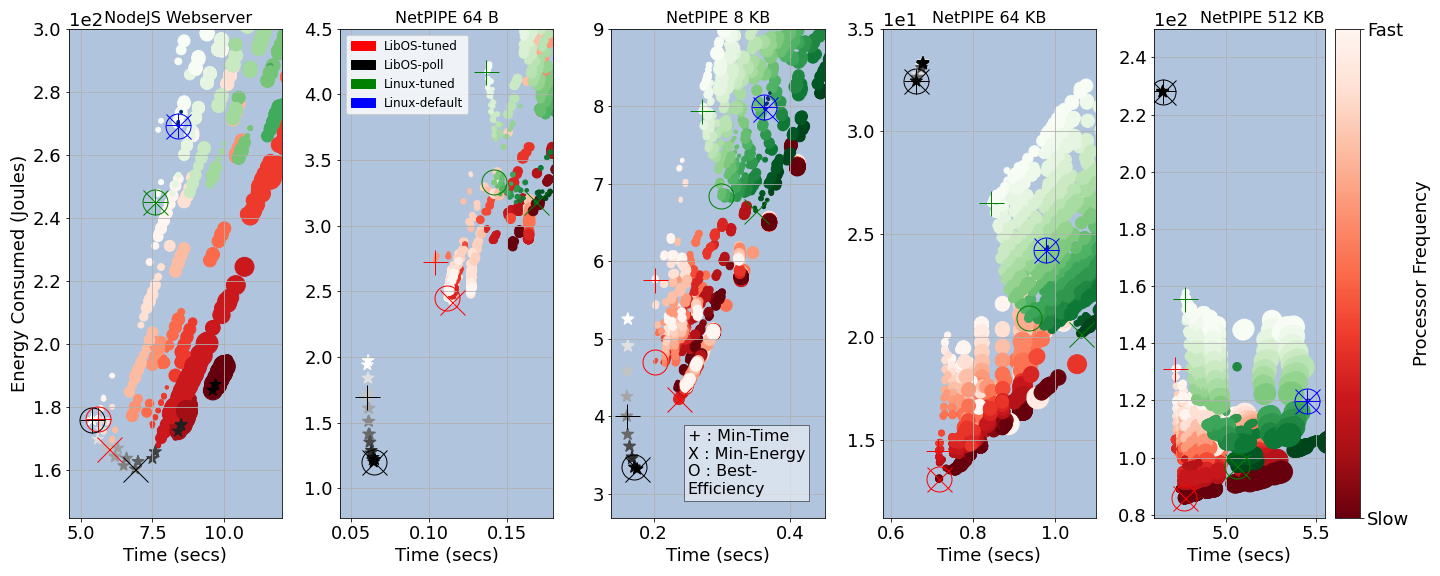}
\caption[]{\small NetPIPE and NodJS Webserver performance and energy results for different message sizes. Every datapoint is the result of a single experimental run with a unique ITR, DVFS combination while \textit{Linux-default} has dynamic ITR-delay, DVFS algorithms enabled instead. LibOS refers to EbbRT. The X-axis is a measure of performance (lower is better) and Y-axis shows the total energy consumed. For \textit{Linux-tuned} (or Linux-static) and \textit{LibOS-tuned} (or EbbRT-static), the labeled (ITR-delay, DVFS) pair are experimental values that resulted in the lowest energy use. \textit{LibOS-poll} shows EbbRT with a run-to-completion polling loop at different processor frequencies (shown as change in gradient colors). \textit{Note: The X and Y scales are different to show the structure of collected data.}}
\label{fig:closed_overview}
\vspace{-0.1in}
\end{figure*}

\subsection{Hardware Mechanisms}
\label{sec:hwmech}
We summarize the software and hardware techniques we use to conduct sweeps of static ITR-delay, and DVFS combinations.

\subsubsection{Interrupt Coalescing (ITR-delay)}
Most modern NICs have a hardware feature to control per-interrupt rates~\cite{intelitr, mellanoxsinterrupt} that induce interrupt coalescing. Typically in Linux, these rates are set dynamically by pre-built dynamic policies within their respective device drivers. However, it is possible to set them statically and we use \textit{ethtool}~\cite{intelethtool} in the Linux study\footnote{\textit{ethtool} is a user tool that maps interrupt coalescing values to appropriate NIC settings}. For EbbRT, we program the NIC directly via its Intel device driver. ITR-delay on Intel NIC's can be programmed in 2 $\mu$s increments.

\subsubsection{CPU Frequency (DVFS)}
DVFS power states (p-states) are features on modern processors that trade-off instruction execution speed for a reduction in energy use~\cite{armdvfs, amdpstate, intel_manual}. Normally, p-states are set dynamically by a policy governor in Linux~\cite{cpufreq_governor}. In this study, we disable dynamic DVFS through Linux's \texttt{userspace} governor and write directly to the IA32\_PERF\_CTL MSR register~\cite{intel_msr} instead. We replicate this in EbbRT by writing to the same register.

\subsection{OS Softwares}
\label{sec:OS}
We explored two OSes with fundamentally different system designs. This gave us the ability to deepen our understanding of how ITR-delay and DVFS mechanisms impact performance and energy consumption under different system implementations.

\subsubsection{Linux}
We build a set of application-specific Linux appliances~\cite{271072, hanappliance} for each of the applications shown in~\cref{table:wrkcfgs}. These appliances are specially constructed to run a RAM-based filesystem and contain only a small set of system libraries and kernel modules required to run their constituent applications. We construct these appliances from a custom 5.5.17 kernel which we built using a modified configuration file for high performance, following suggestions from previous work that studied Linux core operation costs~\cite{linux-core-ops}. To reduce scheduling overheads and noise, we pin all applications to physical cores, disable Linux ~\textit{irqbalance}, and affinitize packet receive interrupts to their respective cores.

\subsubsection{EbbRT}
Specialized systems aimed at accelerating network applications have seen significant research~\cite{ix, arrakis, zygos, shenango, rumpkernel, aliraza, unikernels, scalingmcdfacebook, arachne, mtcp, sandstorm, affinityaccept, flexnic, mica, seda, ebbrt}. However, these systems often overlook importance of their energy efficiency~\cite{SmoothOperator, Dynamo, nature1}. To explore the performance and energy implications of such a specialized system, we chose EbbRT~\cite{ebbrt}, an open-source platform for building per-application library OSes (around 20K LOC). EbbRT shares properties with these prior systems and employs a run-to-completion, event-driven model in a single execution domain. We developed a network device driver for EbbRT for the network applications to run baremetal on servers with Intel 82599 10 GbE NICs~\cite{82599} (around 3K LOC).

\subsection{Per-Interrupt Log Collection}
For the study, we built a per-interrupt logging framework, \texttt{intlog} (Acesss to our data and logging scripts can be found at https://anonymous.4open.science/r/intlog-9925), in Linux and EbbRT's network device driver. We collect the following data in the NIC's interrupt handler code: received and transmitted bytes, descriptors, sleep state statistics, and current timestamp via \texttt{rdtsc} instruction. Additionally, per-core Intel performance monitoring counters (PMCs) capture hardware statistics every millisecond, including instructions, cycles, and last-level cache misses. We use standard Running Average Power Limit (RAPL) hardware registers to read per-package energy values~\cite{intel_rapl}~\footnote{The 1 ms rate is due to sampling granularity of RAPL}. Using rack-level energy measurement mechanisms, we have validated that the changes in energy consumption we observe using RAPL are accurate and impactful\footnote{While we have validated that ITR-delay and DVFS also impact rack-level energy savings, we use RAPL instead because the granularity of the rack-level measurements (on the order of seconds) made it difficult to attribute detailed energy use to specific system events.}.

\subsection{Study Results}
\label{sec:exp}
In our results, we compare and contrast the performance and energy consumption achieved by three OS configurations: 
 \begin{enumerate}
     \item \textbf{Linux}, which has both its dynamic ITR-delay and DVFS algorithms enabled - DVFS is set by Linux \texttt{ondemand} governor~\cite{cpufreq_governor}, while ITR-delay is set by Intel's dynamic policy~\cite{intelitr})
     \item \textbf{Linux-static} and \textbf{EbbRT-static} where ITR-delay and DVFS are set to specific fixed values.
 \end{enumerate}
For both \textbf{*-static} OSes, we conducted a study sweeping to 340 \footnote{This is due to the experimental scope and also to cover a broad range of possible pairs out of 2 million.} static ITR-delay, DVFS pairs, and repeated up to 10 times for stability; our gathered statistics show a standard deviation error of less than 0.01\%. In each experiment, we measure a software stack's performance (elapsed time for closed-loop and 99\% tail latency for open-loop applications) and overall energy usage. While our study generated over 5 TB of data across multiple runs, we will concentrate on presenting three representative findings based on the results of NetPIPE~\cite{snell1996netpipe} and memcached~\cite{memcached} experiments in the following sections.

\paragraph*{Closed Loop} 
NetPIPE is a simple network ping-pong application of fixed-size messages over a single TCP connection and is an example of a closed loop application~\cite{Barroso:2009:DCI:1643608, oldi-study, oldi-pegasus, warehouse-power, energyproportion, WebSearch}. For closed-loop applications, the work to be done is a sequence of requests that have an inter-dependency on each other. Linux runs NetPIPE-3.7.1 while EbbRT uses a custom version ported to its interfaces. 

\paragraph*{Open Loop}
Memcached is an example of an open-loop application, characterized by an external request rate considered largely independent of the time required for request servicing. In our setup, an unloaded client, running mutilate~\cite{mutilate}\footnote{Mutilate is configured to pipeline up to four connections to enhance its request rate.}, interfaces with five agent nodes generating requests to the memcached server. Each agent node operates on a 16-core machine, with each core establishing 16 connections, resulting in a total of 1280 connections. Linux executes memcached-1.6.6, while EbbRT utilizes a re-implemented version tailored to its native interfaces, supporting the standard memcached binary protocol. We run the representative ETC workload from Facebook~\cite{workloadanalysisfacebook}. It uses 20 to 70-byte keys and 1-byte to 1-KB values and contains 75\% GET requests. We use a stringent SLA objective where the 99\% tail latency < 500 $\mu$s.\\

\begin{figure}[h!]
\centering
\includegraphics[width=0.45\textwidth]{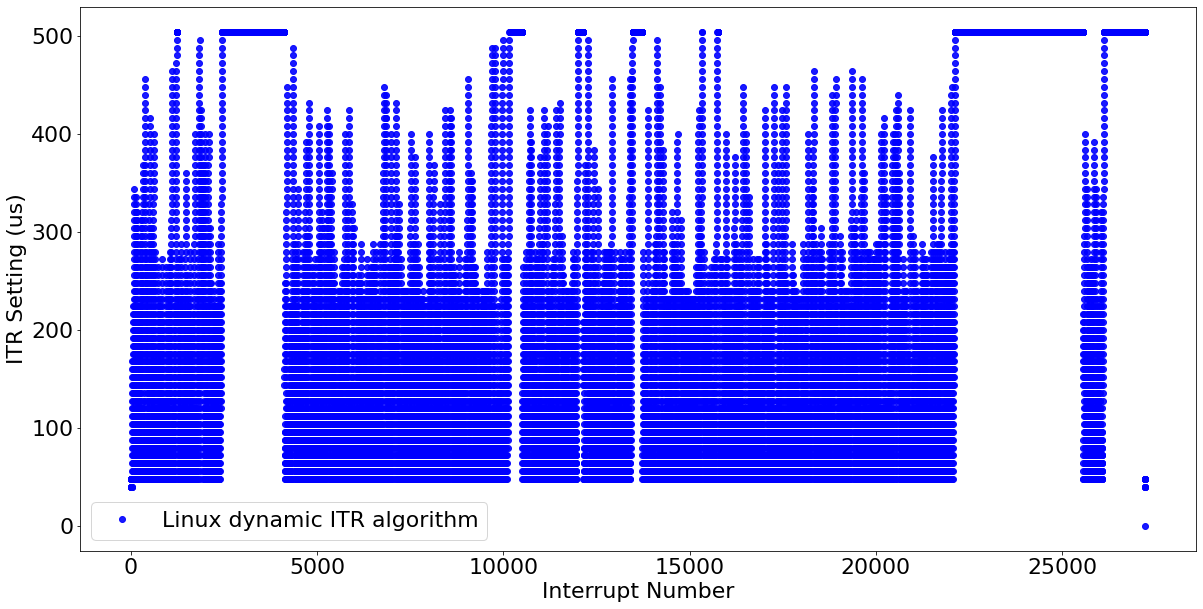}
\caption[]{\small ITR-delay values set by Linux's dynamic ITR-delay algorithm. This is captured during a live run of NetPIPE at 64 KB message size.}
\label{fig:netpipe_65536_itr_vals}
\end{figure}

\subsubsection{ITR-delay and DVFS impact on batched packet processing}
\label{sec:closed_energy}
While the focus of our work is on open-loop style SLA-driven applications, we begin our study with a NetPIPE server. NetPIPE's closed-loop style and simple application protocol allow us to explore how different message sizes, ITR-delay and DVFS affect the overall performance and energy of different OSes. Fig.~\ref{fig:closed_overview} shows that at 64 KB message size, the static ITR-delay in Linux demonstrates a performance improvement of over 60\%, and a 50\% reduction in energy consumption compared to dynamic policies in Linux. To understand why this is dramatic, consider the dynamic ITR-delay policy, visualized in fig.~\ref{fig:netpipe_65536_itr_vals}, which reveals extreme variability at a per-interrupt granularity\footnote{We instrumented a simple log in Linux's network device driver to save every updated ITR-delay value during a single run of NetPIPE for a 64 KB message.}. This indicates that the current policy, designed for general use cases, operates at an inappropriate timescale for NetPIPE and that significant advantages can be gained through specialization. Moreover, fig.\ref{fig:closed_overview} illustrates the Pareto-optimal performance-energy curve for various message sizes in both Linux and EbbRT. As the NetPIPE message size increases from 8KB to 64KB and 512KB, the fixed ITR-delay values yielding optimal energy efficiency also increase toward 26$\mu$s and 28$\mu$s at 512KB for EbbRT and Linux, respectively (labeled in red and green boxes). Intuitively, this result indicates that ITR-delay effectively batches processing by controlling the amount of payload transmitted from the NIC to the OS within a time window. Optimal ITR-delay settings suggest a "sweet spot" where the OS paces packet processing, saving energy through a combination of idling and CPU frequency control.

Lastly, though (ITR-delay, DVFS) pairs in fig.~\ref{fig:closed_overview} have different values for the different OSes explored, the performance-energy curves for the OSes follow a common 'V' shape. The lowest point in this 'V' shape represents a setting that consumed the least energy while being competitive in performance while the left points represent settings that sacrificed energy for better performance. This 'V' shape also illustrates that it is essential to be strategic in tuning, as while some static settings can outperform dynamic control, there can also be sub-optimal static settings, as shown by points to the right of Linux in Fig.~\ref{fig:closed_overview}. 

\begin{figure*}[!htb]
\centering
\includegraphics[width=0.99\textwidth]{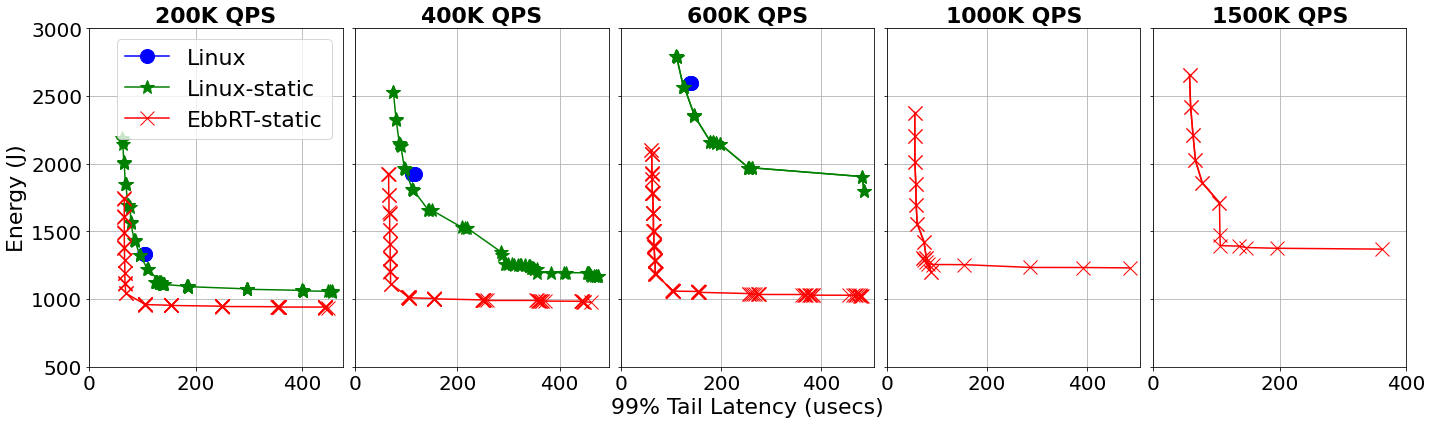}
\caption[]
{\small Memcached: Each point represents a single experimental run. The \textit{*-static} data points use a unique (ITR-delay, DVFS) pair. We only illustrate data that lie on the Pareto-optimal curve. The X-axis shows performance measurement (lower is better) and the Y-axis shows total energy consumed.
\textit{Linux results for 1000K and 1500K QPS loads are not shown as Linux could not support them without violating SLA.}
}
\label{fig:mcd_overview}
\end{figure*}

\begin{figure}[!htb]
\includegraphics[width=1\columnwidth]{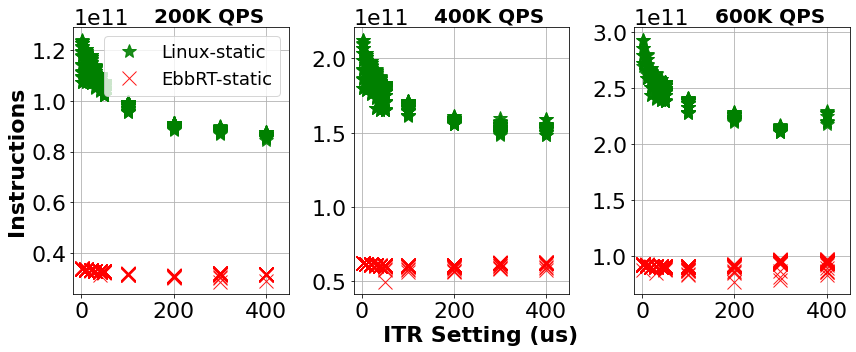}
\caption[]{\small Memcached: ITR-delay impact on instruction count ($1e11$). 
\textit{Not drawn to scale to show structure in data.}
}
\label{fig:sig_mcd_ins_itr}
\end{figure}

\subsubsection{OS Specialization on Energy and Performance}
\label{sec:open1}
Next, we consider experiments that explore the performance and energy trade-offs of memcached with varying requests-per-second (QPS) loads under the same SLA objective. Our key findings are that 1) different OS designs can impose different trade-offs between performance and energy, and 2) specialized systems can achieve dramatic efficiency in both. This can be seen in fig.~\ref{fig:mcd_overview} which illustrates the Pareto-optimal curves \footnote{We filter out (ITR-delay, DVFS) pairs that resulted in SLA violations.} of EbbRT and Linux. Fig.~\ref{fig:mcd_overview} shows that as QPS increases, EbbRT exhibits a consistent vertical structure, suggesting effective energy savings without impacting latency. Conversely, Linux's curves become more horizontal, indicating performance degradation as QPS rises due to increased trade-offs between performance and energy. Notably, EbbRT's optimized stack allows it to handle higher peak QPS (2000K) compared to Linux (800K). In particular, fig.~\ref{fig:sig_mcd_ins_itr} shows the impact of ITR-delay on the total amount of instructions needed to run a single memcached server in both Ebbrt and Linux. This figure shows how a large ITR-delay (e.g. 400 $\mu$s) can reduce overall instruction count by up to 30\% in Linux. It also shows the drastic differences in instruction count between the two OSes, as EbbRT uses up to 2.5X fewer instructions to support the same load as Linux does. This implies that a greater fraction of EbbRT's instructions were spent getting actual work done rather than traversing the network stack, which suggests that combining ITR-delay and DVFS control with EbbRT's optimized network paths presents substantial opportunities for maximizing \textit{race-to-idle} energy benefits~\cite{dreamweaver, dynsleep}.

\begin{figure}[!htb]
\begin{subfigure}{\textwidth}
  \includegraphics[width=0.5\linewidth]{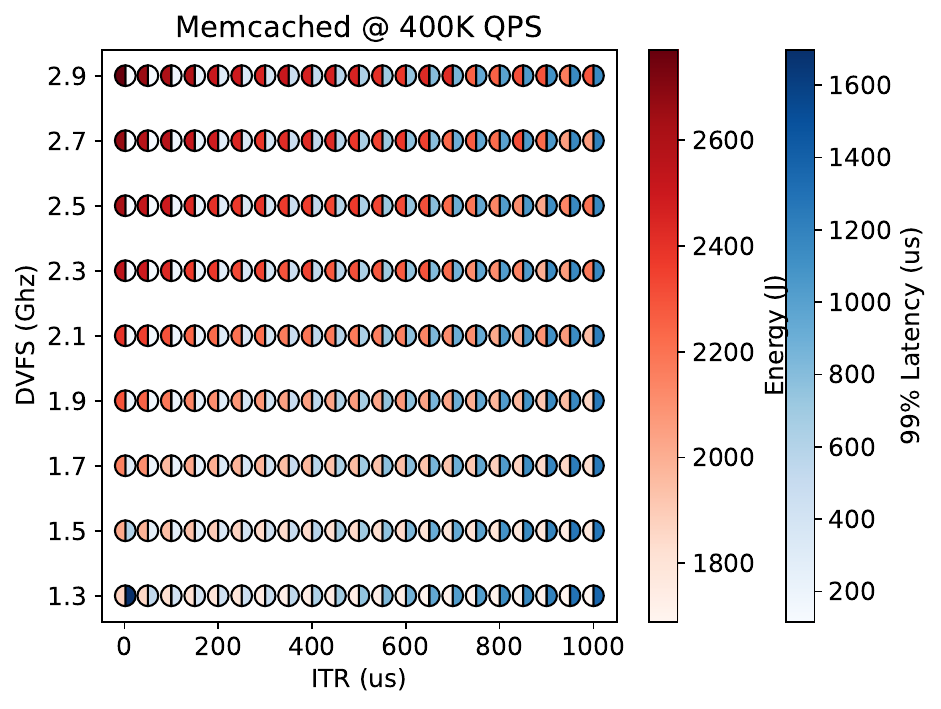}
\end{subfigure}
\caption{\small Illustrates the change in energy and 99\% latency as different ITR-delay, DVFS pairs are explored for Linux memcached.}
\label{fig:gradients}
\end{figure}

\subsubsection{Polling can be energy efficient}
As indicated earlier, the use of a single fast ITR value resulted in both best performance and subsequent combining with DVFS also resulted in lowest energy use as well. This prompted us to explore the effects of eliminating interrupts altogether and use a dedicated poll loop. In \cref{fig:closed_overview}, we illustrate that LibOS-poll (or EbbRT-poll) was able to achieve both best case latency and competitive lowest energy. We found that by modulating DVFS, EbbRT-poll can be made energy efficient for small payloads under its specialized OS paths - this is in contrast to the normative assumptions of OS poll where it often trades performance for higher energy use. For example, EbbRT was able to improve tail latency by 27\% while using 35\% less energy in Memcached. In NetPIPE~\cite{snell1996netpipe}, polling can achieve up to 3X better performance while using 4X lower energy as compared to baseline Linux. However, polling with DVFS must be used judiciously as in other application-centric workloads such as Memcached-silo often results in both worst performance and energy use as compared to their interrupt-driven counterparts.

This finding suggests the importance for energy aware OS-level optimizations that can switch between poll and interrupt-driven IO in response to changes in demand. Therefore, OS path specialization techniques can explore the use of polling to achieve both low-latency and energy efficiency with careful use of DVFS in new hybrid policies.

\subsubsection{Revealing ITR-delay and DVFS Performance-Energy Trade-offs}
\label{sec:open2}
To help build intuition of the impact of specific ITR-delay and DVFS settings on the performance and energy of open-loop style applications, we present an example in \cref{fig:gradients} featuring a Linux memcached server with a load of 400K QPS. Using colored gradients, the figure visually represents the effects of each ITR-delay, and DVFS pair; each data point is divided in half, providing insights into their respective impacts on 99\% latency and energy.

In \cref{fig:gradients}, one can see the trend that as DVFS decreases from 2.9 GHz: the energy gradient becomes lighter, indicating a more pronounced impact on reducing energy use. Simultaneously, increasing ITR horizontally has an immediate effect on increasing measured 99\% latency, evident in the darkening of colored gradients. Further, we observe two notable behaviors at a DVFS frequency of 1.3 Ghz: 1) a fast ITR-delay (0 $\mu$s) triggers a spike in tail latency, violating the 500 $\mu$s SLA objective due to the slow CPU frequency's insufficient processing of incoming requests, and 2) as ITR-delay increases, this induces additional queuing which enables efficient request batching, thereby facilitating additional energy savings. 

These observations indicate that the combination of ITR-delay and DVFS enables one to select different operating points that can move within this space. This is evident in the common "L" shapes seen in \cref{fig:mcd_overview}; which while they differ in absolute performance and energy, illustrates how ITR-delay and DVFS can be combined to reduce energy while still meeting SLA objectives in both OSes.

%% file: model2.tex
\section{Modeling ITR-delay and DVFS Effects}
\label{sec:model}
A key takeaway from \cref{sec:study} is that \cref{fig:closed_overview} and \cref{fig:mcd_overview} reveal common shapes ("V" and "L") that are OS-agnostic and share a stable structure in response to changes to ITR-delay and DVFS. This suggests one can develop a formal model that captures OS-agnostic performance and energy profiles and that generic external control mechanisms can then be made feasible for both OSes.

\subsection{Memcached Model Fitting}
Motivated by the implications of these OS shapes, we formulated a mathematical model to explore fitting our experimental data with a set of free parameters and ITR-delay, DVFS settings. We assume a simple model where the offered load is light enough that requests don't batch up in the receive queue and can be treated independently. We model the performance as 99\% tail latency as well as energy consumed \footnote{We have also collected other tail latency values and found that our model can accurately fit them as well.}. 

\subsubsection{Performance:} We define $\triangle t$ as the time it takes to handle a single request:
$$ \triangle t = t_{work} + t_{interrupt} $$

We parameterize $t_{work}$ as a function of DVFS values:

\begin{equation} \label{eq:t_work}
t_{\text{work}} = \frac{Z}{DVFS^{1+\alpha}}
\end{equation}

$Z$ and $\alpha$ are free parameters that change for both the OS and application load. In this model, $Z$ acts as a maximum time limit that each request can take (i.e. SLA objective). $\alpha$ represents a system's dependence on DVFS to trade off performance for energy. For example, if $\alpha = -1$, then that particular system has no dependence on DVFS and can largely use DVFS to lower energy use without sacrificing performance - this is inspired by the study results in \cref{sec:open1} that illustrate how DVFS affects Linux and EbbRT differently in trading off energy for latency.

We parameterize $t_{interrupt}$ as a function of ITR-delay values:
\begin{equation} \label{eq:t_interrupt}
    t_{\text{interrupt}} = \phi * ITRdelay
\end{equation}

As ITR-delay greatly affects the measured tail latency, $\phi$ represents the location in the receive queue where a packet is placed before being processed. For example, if an unlucky packet arrives just as the NIC's $ITR$ value starts counting down, then it will have to artificially wait for a full $ITR$ before being processed, thereby delaying overall request processing time. 

\subsubsection{Energy:} We define $\triangle J$ as the amount of energy it takes to process a single request. This is affected by the voltage and frequency states of DVFS and how ITR-delay can induce prolonged idle periods: 

\begin{equation} \label{eq:open_energy}
\triangle J = \gamma*(\phi*ITR)*DVFS^\beta 
\end{equation}

Note that $\phi$ used here is the same variable from \cref{eq:t_interrupt}. $\gamma$ (units of watts) acts to convert the interactions of ITR-delay and DVFS into energy used. The variable $\beta$ acts as a dependence factor on DVFS in a similar way to $\alpha$ in \cref{eq:t_work}.

\begin{figure}
\centering
    \includegraphics[width=0.48\textwidth]{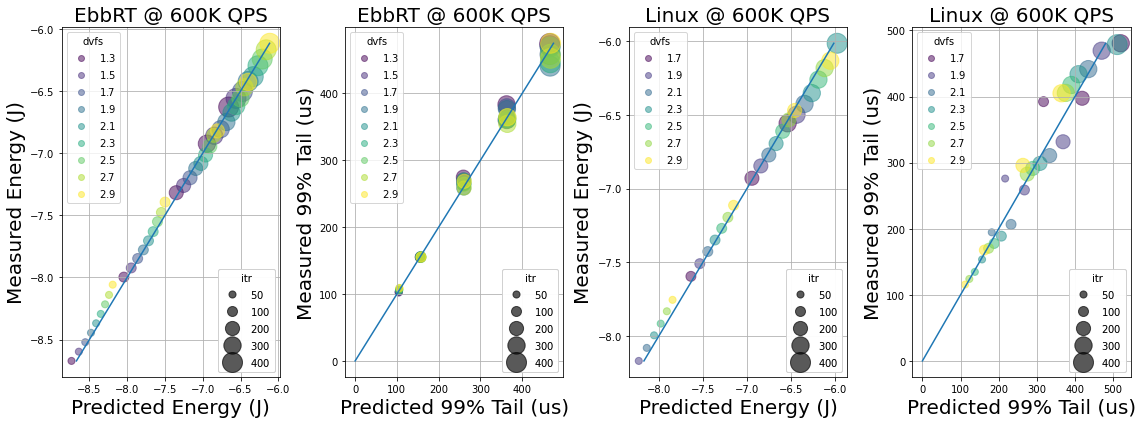}
    \caption{\small Prediction of energy and performance across both OSes using our model for Memcached @ 600K QPS. The diagonal line indicates a perfect model fit. The negative energy values are due to \texttt{log()} transformations during modeling for regression analysis.}
    \label{fig:mcd_model_600K}
    \vspace{-0.2in}
\end{figure}

Fig.~\ref{fig:mcd_model_600K} illustrates one example result of the model fitting against memcached data for a QPS of 600K. The x-axis shows the set of energy and performance predictions and the y-axis shows their measured values. The diagonal lines show where ideal points would lie if our model's calculations were exact. We use the Adam optimizer from PyTorch~\cite{pytorchadam} in this process and run each fit several times to check the stability of the inferred parameters and avoid getting stuck in local minima. Overall, we find that despite complicated interactions between the hardware and software, our models are expressive enough to exploit the common OS response behaviors to accurately fit both performance and energy data. 

\begin{figure*}[htb!]
\centering
\includegraphics[width=0.85\textwidth]{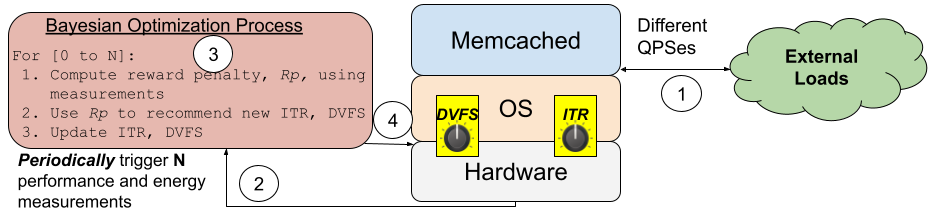}
\caption[]
{\small Our controller designed to optimize energy efficiency for a memcached server.}
\label{fig:bayop}
\vspace{-0.15in}
\end{figure*}

However, the limitation of our model is that it is only practical in highly constrained settings. To replicate this approach for new software and hardware, one would need to exhaustively re-gather data while tuning ITR-delay and DVFS. Nevertheless, the accuracy of our model suggests the viability of using similar approaches to search this space.

%% file: bayop.tex
\section{Proof-of-Concept Controller}
\label{sec:cachetrace}
Motivated by our prior findings and modeling work, we present an example controller to help validate our conjecture of a black-box search strategy. This controller uses an established machine learning technique, Bayesian optimization~\cite{frazier, garnett_bayesoptbook_2022}, to find energy-efficient interrupt coalescing and CPU frequency settings that can adapt to the specific application, OS, and hardware while exploiting the stability in offered loads. 

In 1) \cref{sec:cachetrace2}, we illustrate its applicability in optimizing the energy efficiency of a server that supports a realistic datacenter workload trace~\cite{cacheWorkload-OSDI20} over 24 hours by periodically adjusting ITR-delay, and DVFS settings as offered load changes, and in 2) \cref{sec:tailbench} demonstrates the generality of the controller as we apply it across different types of NICs and CPUs (\cref{table:hwsw}) when run on three applications from Tailbench~\cite{tailbench}.

As a proof-of-concept controller, we made simplfying assumptions in its design and leave addressing real deployment scenarios for future work. Our assumptions include ad-hoc thresholds for when to trigger Bayesian optimization and the number of subsequent trials to run. However, our results show that even using straightforward assumptions can yield significant advantages, leaving ample room for improvement.

The architecture of our controller also facilitates the integration of more advanced policies for initiating the Bayesian process. We envision the deployment of this technique in data centers through collaboration with load-balancers that make use of historical usage data. This collaboration would help distribute incoming loads to energy-optimized servers, which have been pre-configured with specific settings, while still meeting SLA objectives. In addition, the load balancers can help mitigate the potential gaming of the learning agent's behavior in response to changing request rates.

\subsection{Design}
Fig.~\ref{fig:bayop} illustrates the design of our  controller: \circled{1} A live system running memcached services requests arriving at varying QPSes from an external source. \circled{2} It then triggers a set of performance and energy measurements of the live system to be shared with an external Ax~\cite{ax, Bakshy2018AEAD} Bayesian optimization platform. \circled{3} This process then computes a penalty $Rp$ of the current (ITR-delay, DVFS) setting and \circled{4} then recommends an update to a new (ITR-delay, DVFS) configuration on the live system that minimizes $Rp$. Once this process completes, the live system is set with a fixed (ITR-delay, DVFS) configuration until the next set of measurements is triggered.

\subsubsection{Penalty Function}
We use a simple function that penalizes the optimization process by the amount of measured energy and magnifies that penalty when measured latency violates the SLA objective:

\begin{equation} \label{eq:bayesopt_reward}
    Rp = m\_energy * max((m\_latency - SLA + 1), 1)
\end{equation}

For example, with an SLA objective of 99\% tail latency < 500 $\mu$s, where measured latency ($m\_latency$) is 600 $\mu$s, the reward $Rp$ will be scaled up by a factor of 100, such that $Rp = m\_energy * (600 - 500)$. If $m\_latency$ is less than $SLA$, then $Rp$ will evaluate to $m\_energy$. Minimizing $Rp$ is indicative that Bayesian optimization is selecting (ITR-delay, DVFS) pairs to meet performance/energy objectives. This reward function enables an operator to express their preference to optimizing for different combinations of energy and performance objectives.

The possibilities of customizing this function further are also ripe for exploration: such as using new combinations of performance/energy or known metrics such as energy-delay-product~\cite{573184,10.1109/40.888701}. One can also imagine developing a rich set of reward functions that capture preferences a service operator might have. In this way, the controller can be reconfigured as priorities change by selecting and tuning from the set of reward functions.
\input{cachetrace3}

\input{tailbench}

%% file: cachetrace3.tex
\subsection{Applicability to \textit{cache-trace}}
\label{sec:cachetrace2}
This section presents the results of running our controller against a publicly available KV store workload trace (\textit{cache-trace}~\cite{cacheWorkload-OSDI20}) which exhibits the stable demand curve behavior for our controller. 

\subsubsection{Experimental Setup}
We used the same infrastructure of our study (\cref{sec:study}) but modified the \textit{mutilate} workload generator to generate QPSes following from \textit{cache-trace} instead: first, we extracted a 24-hour sequence of QPS rates from a single trace and binned the data into hourly divisions to capture the mean QPS rate at an hourly basis. As cache-trace QPS rates were often in the tens of thousands of QPS as it was running on limited vCPUs, we scaled up the rates to match our hardware capability. However, \ref{sec:open2} shows that even at low QPS rates where DVFS is fixed at the lowest CPU frequency, ITR-delay can still be used to further reduce energy use. Therefore, we then generate these scaled-up mean QPSes to our live memcached server for which we capture energy-per-second measurements over the entire 24-hour period. 

The controller is configured to trigger its periodic measurements at an hourly rate and run Bayesian optimization for a default of 30 trials - this is due to overheads in our single-thread Python package; which takes around 5 minutes to run. In contrast to our initial energy study (\cref{sec:study}), which was limited to only using up to 340 (ITR-delay,  DVFS) pairs due to experimental scope, our controller allows Bayesian optimization to choose from all available ITR-delay, and DVFS values (a total of \textit{2 million} possible combinations). 

We evaluate our controller by comparing the energy and performance behavior of five different system configurations:
\begin{itemize}
    \item \textbf{Linux}: Operating in its default state, where the dynamic ITR-delay and DVFS  algorithms are enabled.
    \item \textbf{Linux-BayOp and EbbRT-BayOp}: Operating with Bayesian Optimization to tune both ITR-delay and DVFS, with a target of minimizing overall energy use while maintaining SLA objectives.
    \item \textbf{Linux-DVFS-BayOp and Linux-ITR-BayOp}: Operating with Bayesian Optimization to tune only one of the two settings. We were motivated to explore these configurations to better understand the limitations of the two hardware mechanisms individually. \textit{Linux-DVFS-BayOp} tunes DVFS while enabling the dynamic ITR-delay algorithm. \textit{Linux-ITR-BayOp} tunes ITR-delay while enabling the dynamic DVFS algorithm.
\end{itemize}

\subsubsection{Evaluation}
We evaluate our controller's energy impact across two
applications, namely memcached and silo, in both Linux and EbbRT\footnote{The controller's penalty can also be modified to minimize latency, details can be found in \cref{sec:appendix}}. Silo~\cite{mcdsilo, zygos} is a compute and memory-intensive application that is extended with a web front-end such that every request triggers a corresponding set of TPC-C transactions on an in-memory database~\cite{silo}. We ported Silo to EbbRT and the workload mix and SLA constraints of Silo follow from those used in memcached.

\begin{figure}[!htb]
\centering
\begin{subfigure}{.45\textwidth}
  \centering
  \includegraphics[width=\linewidth]{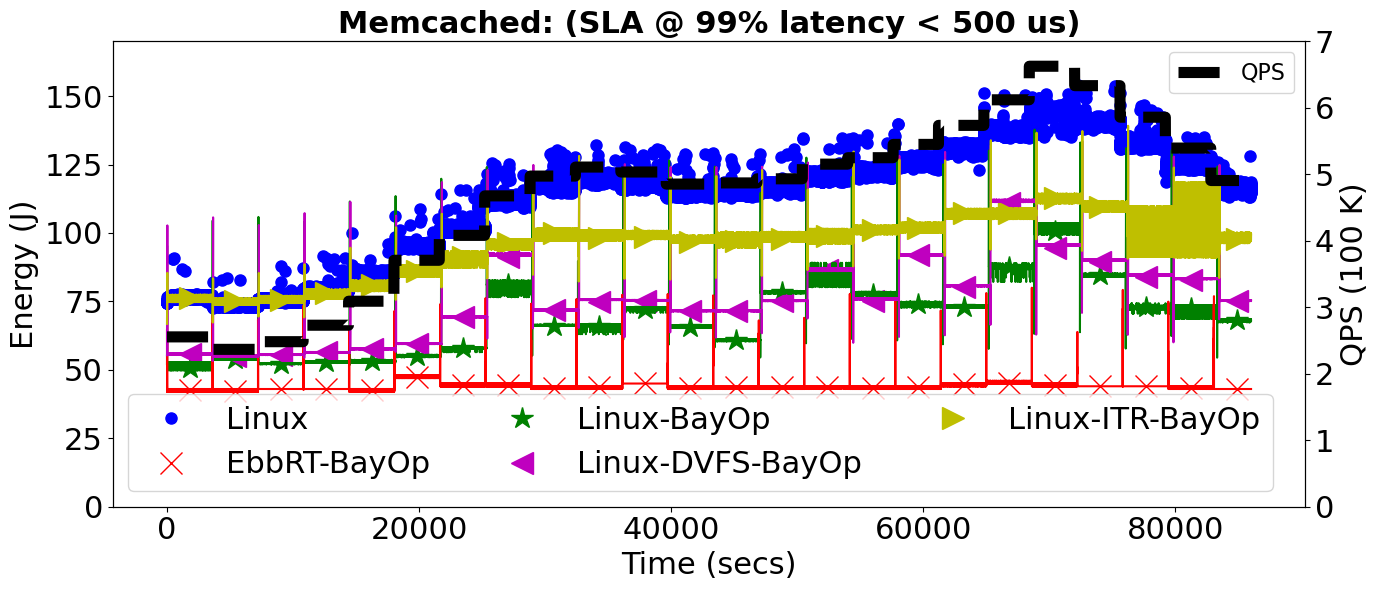}
\end{subfigure}
\begin{subfigure}{.45\textwidth}
  \centering
  \includegraphics[width=\linewidth]{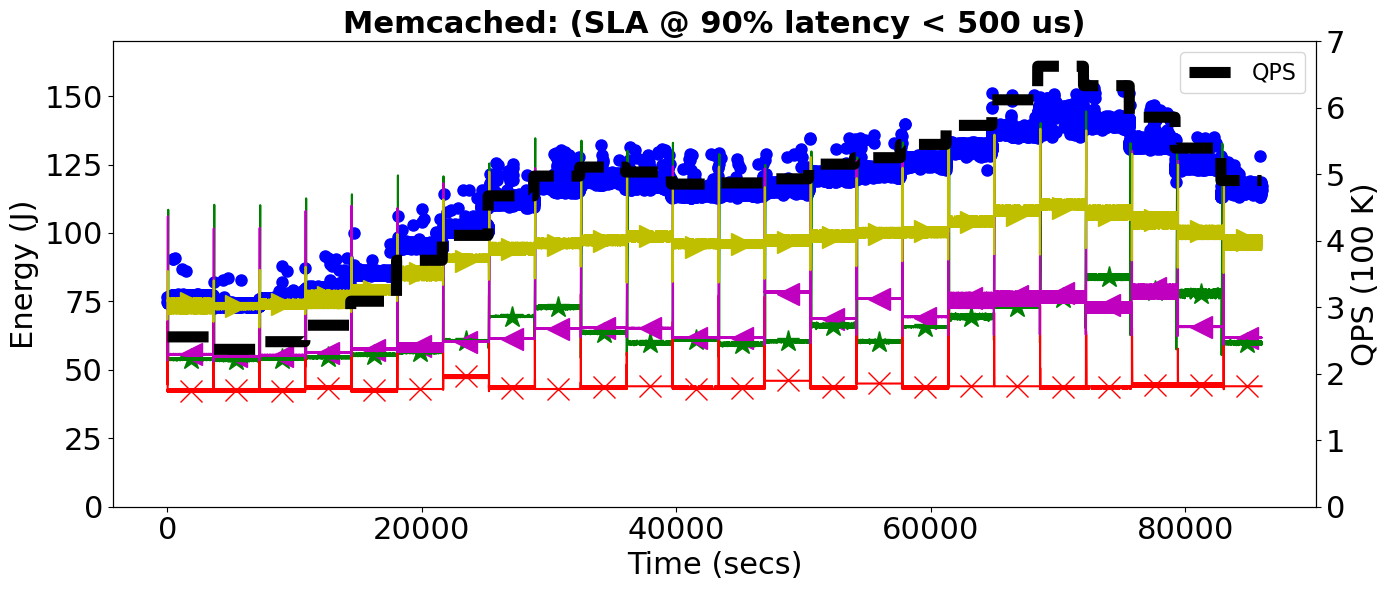}
\end{subfigure}
\begin{subfigure}{.45\textwidth}
  \centering
  \includegraphics[width=\linewidth]{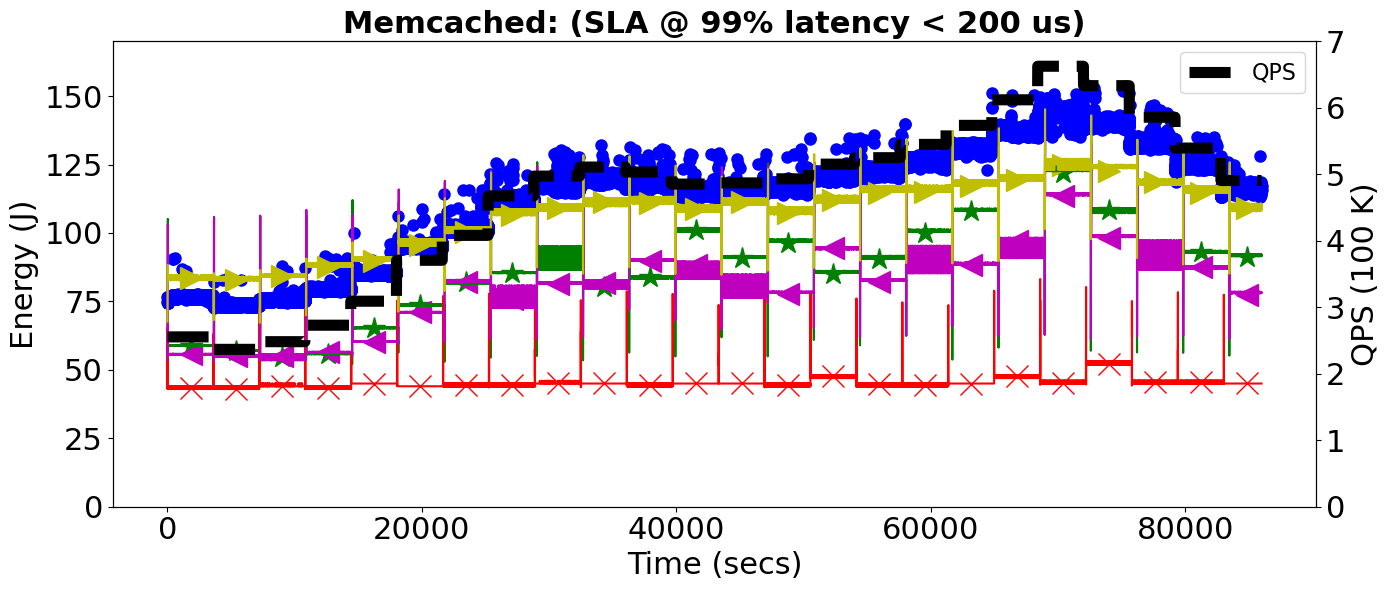}
\end{subfigure}%
\caption{\small \texttt{BayOp} applied to memcached; the \textbf{QPS} label is shown on the right side of the graph and the QPS lines show the different offered loads on a per-hour basis. We present results from different SLA objectives studied and illustrate the measured power (energy/second) on the Y-axis as QPS changes across the five system configurations studied over 24 hours (X-axis).}
\label{fig:mcd_bayop}
\vspace{-0.1in}
\end{figure}

\paragraph{Memcached Results}
Fig.~\ref{fig:mcd_bayop} illustrates our controllers evaluation against three different SLA objectives: 99\% latency < 500 $\mu$s, 90\% latency < 500 $\mu$s, and a even more stringent 99\% latency < 200 $\mu$s. The QPS values, shown on the right, change on an hourly basis, as shown by black line segments. At the beginning of each hourly QPS change, we see spikes in energy usage of \textit{*-BayOp} systems which results from the Bayesian Optimization process searching through (ITR-delay,  DVFS) settings on the memcached server to meet its optimization objective. After this initial energy spike, the system settles into a steady energy consumption state until the next hourly trigger. A key result of this application is the importance of using both ITR-delay and DVFS to meet SLA objectives for optimizing energy efficiency rather than individually.

\begin{figure*}[ht!]
\centering
\begin{subfigure}{.49\textwidth}
  \includegraphics[width=\linewidth]{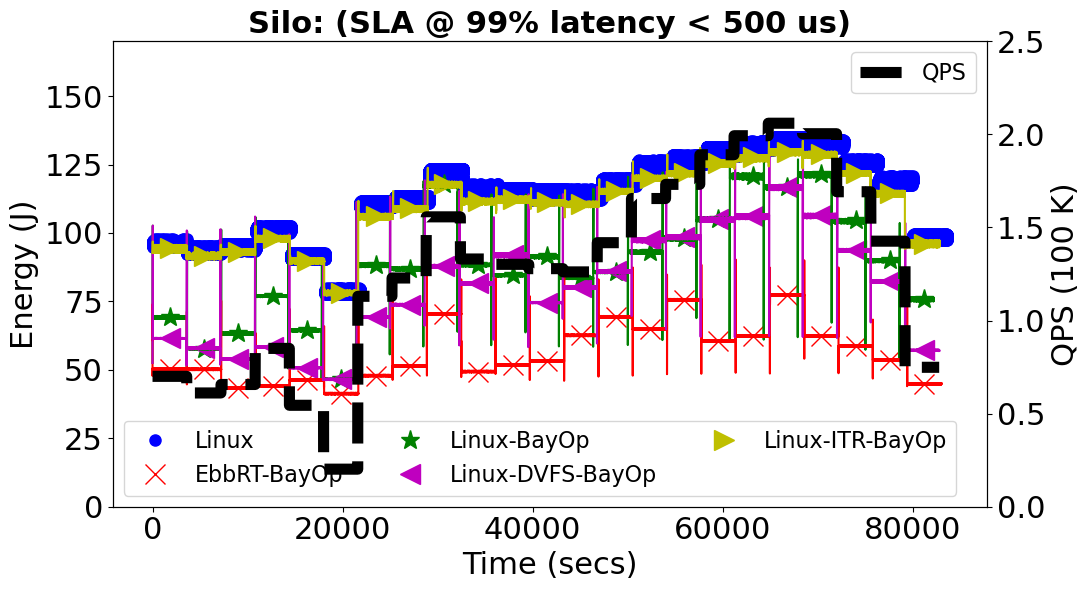}
\end{subfigure}
\begin{subfigure}{.49\textwidth}
  \includegraphics[width=\linewidth]{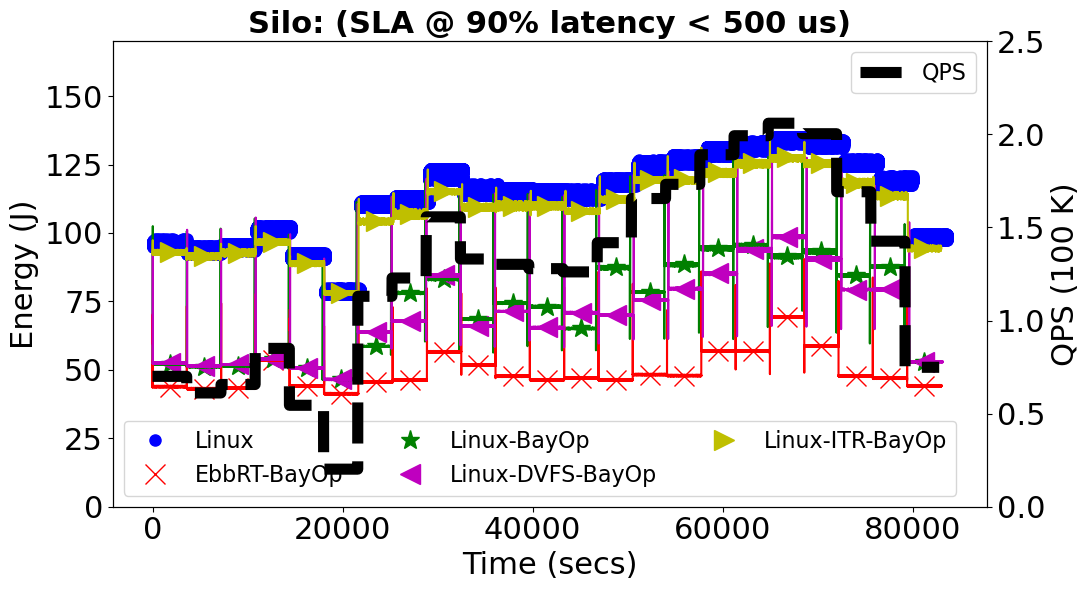}
\end{subfigure}
\caption{\small Controller applied to \textit{cache-trace} for Silo. We show two different SLA objectives. The \textbf{QPS} line shows the change in QPS offered load on a per-hour basis. The consumed power (energy/second) of each system configuration on the Y-axis is shown over 24 hours on the X-axis.}
\label{fig:mcdsilo_bayop}
\vspace{-0.15in}
\end{figure*}

\begin{figure}[h!]
\centering
    \includegraphics[width=0.45\textwidth]{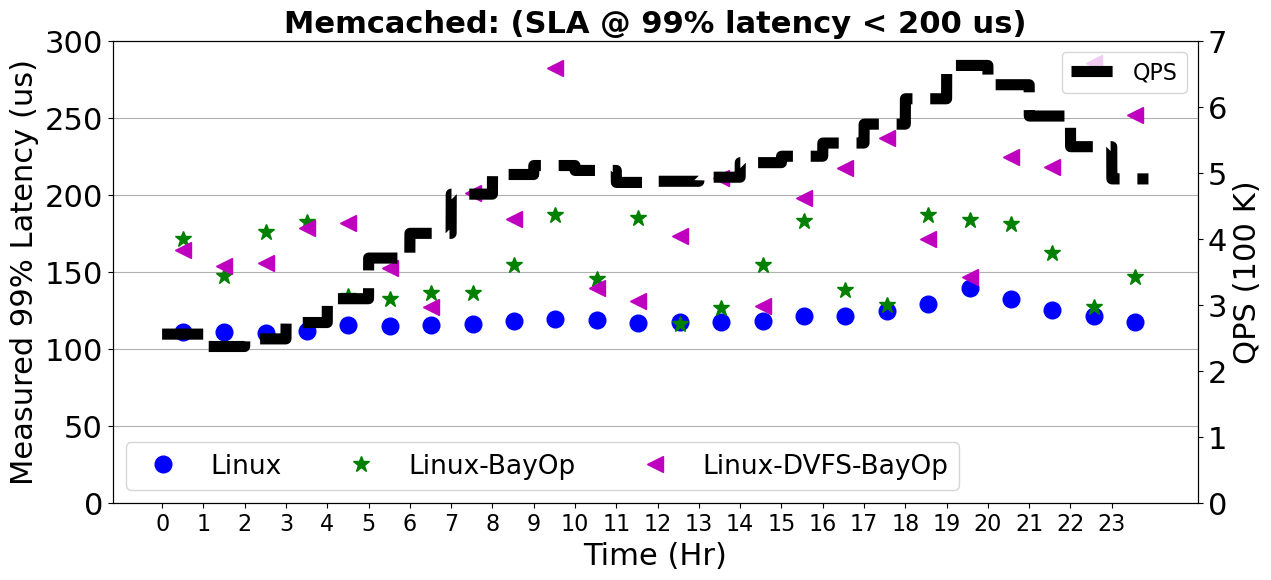}
    \caption{\small Measured 99\% latency across Linux for an SLA of 200 $\mu$s. The latency is shown on a per-hour basis due to how \textit{mutilate} reports its resultant latency measurements. We find that \textit{Linux-DVFS-BayOp} often violates the SLA which suggests only tuning DVFS is not enough to achieve stable system behavior.}
    \label{fig:bayesopt_mcd_lat200_per99_lat}
    \vspace{-0.25in}
\end{figure}

We find that, for an SLA objective of 99\% latency < 500 $\mu$s, \textit{Linux-BayOp} can result in energy savings of up to 50\% over \textit{Linux}. Relaxing the SLA objective to 90\% < 500 $\mu$s enables our controller to find (ITR-delay,  DVFS) configurations that yield even more energy savings of over 60\%. At the most stringent SLA of 99\% latency < 200 $\mu$s, our controller can still adapt while enabling energy savings of up to 30\%. 
The energy savings of \textit{EbbRT-BayOp} are similar to those found in our energy study of memcached (Fig.~\ref{fig:mcd_overview}). Our controller is robust enough to adapt to the software stack of EbbRT and find energy-efficient configurations that consistently result in the lowest energy use (over 2X lower than \textit{Linux}). The measured energy-per-second variability of EbbRT is often lower in contrast to that of Linux (indicated by the thinner red plot in Fig.~\ref{fig:mcd_bayop}), a byproduct of EbbRT's simplified and more optimized network paths.

For \textit{Linux-ITR-BayOp}, allowing our controller to tune only ITR-delay still generally improved energy savings over \textit{Linux}. However, for a stringent SLA of 99\% latency < 200 $\mu$s, the reduced SLA headroom prevents the controller from trading off latency for energy as effectively as it can when tuning alongside DVFS. At the lower QPSes, \textit{Linux-ITR-BayOp} performed worse than \textit{Linux}.

Allowing our controller to tune only DVFS (\textit{Linux-DVFS-BayOp}) results in energy savings comparable with \textit{Linux-BayOp} across SLA objectives. This is further supported by \ref{sec:open2} which illustrates the significant influence of DVFS on overall energy consumption. However, though it may seem that under a more stringent SLA of 99\% latency < 200 $\mu$s, \textit{Linux-DVFS-BayOp} results in the highest energy savings, we found instances where the measured 99\% latency violated the SLA of 200 $\mu$s, as shown in Fig.~\ref{fig:bayesopt_mcd_lat200_per99_lat}; revealing the weakness of relying on DVFS only.

\begin{figure*}[!htb]
\begin{subfigure}{\textwidth}
  \includegraphics[width=\linewidth]{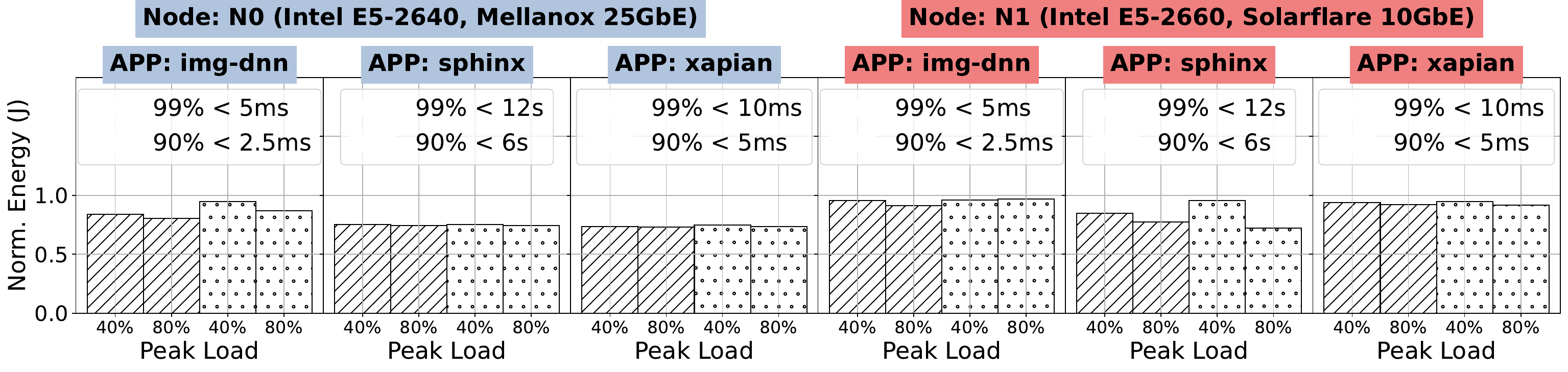}
\end{subfigure}
\begin{subfigure}{\textwidth}
  \includegraphics[width=\linewidth]{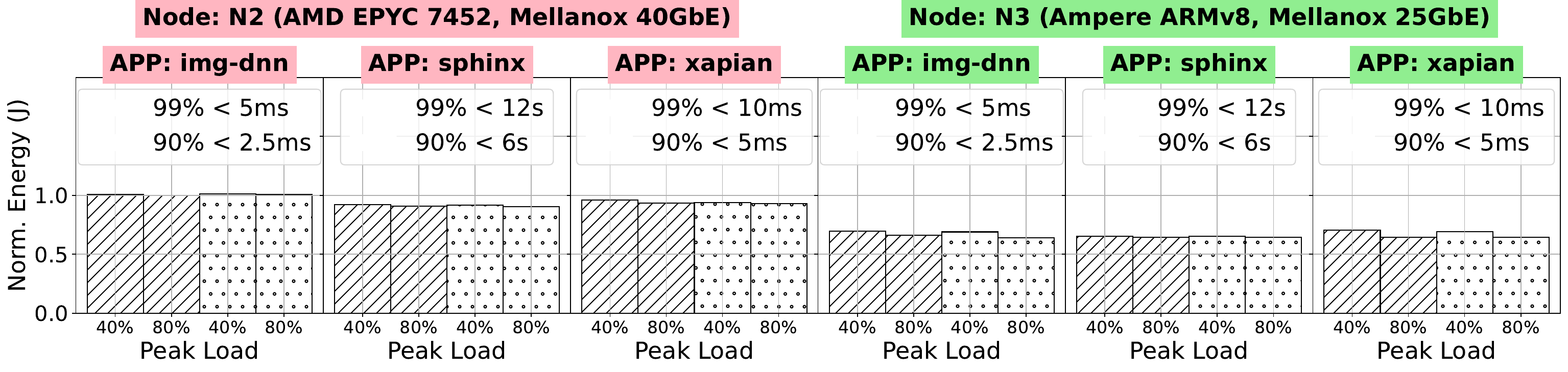}
\end{subfigure}
\caption{\small This figure illustrates the energy use of each application (\textbf{APP}) for each of the hardware platforms (\textbf{NODE: N0 to N3}). The energy is normalized (Y-axis) against Linux default, where lower is better. For each \textbf{APP}, we use two representative offered loads which are 40\% and 80\% of the measured \textbf{Peak Load} of Linux default. Within each representative offered load, we also selected two \textbf{SLAs} (as indicated by \textbf{/////} and \textbf{.....}) for the application to meet while our controller is optimizing its energy efficiency.}
\label{fig:xl170}
\vspace{-0.1in}
\end{figure*}
\paragraph{Silo Results}
We selected another trace from \textit{cache-trace} that was akin to a more computationally intense server. Fig.~\ref{fig:mcdsilo_bayop} shows that the trace peak QPS rates are often lower than those of Fig.~\ref{fig:mcd_bayop} (peak 250K QPS versus 750K QPS). Fig.~\ref{fig:mcdsilo_bayop} does not show results for SLA of 99\% latency < 200 $\mu$s, as the inherent computational cost of Silo's TPC-C transactions resulted in a lower bound of measured latency values that were consistently greater than the SLA objective of 200 $\mu$s. A key result of this application is that it helps expose in computationally intensive cases the limitation of ITR-delay to affect energy savings.

Fig.~\ref{fig:mcdsilo_bayop} illustrates that even for a computationally intensive application with different SLA objectives, \textit{Linux-BayOp} was able to find (ITR-delay,  DVFS) settings that enable 30\% energy savings in Linux for various QPS rates and higher winnings when the SLA is relaxed to 90\% latency < 500 $\mu$s.

The controller was able to adapt to a different OS and application stack and found configurations of \textit{EbbRT-BayOp} that consistently had the lowest energy use over Linux. In contrast to Fig.~\ref{fig:mcd_bayop}, one can see larger variations in energy saved from one QPS to the next (more hilly behavior). This can be partly attributed to the complicated database work that must now be done per request.

In contrast to memcached, we find that tuning ITR-delay alone (\textit{Linux-ITR-BayOp}) while enabling Linux's default DVFS mechanism is largely ineffective at reducing energy. This is likely due to the increased computational cost for each request which limits the potential energy savings gained from interrupt coalescing and prolonged sleep states that are induced by the ITR-delay mechanism.

We find that tuning DVFS alone (\textit{Linux-DVFS-BayOp}) while enabling Linux's default ITR-delay mechanism works surprisingly well for Silo and, in most cases, achieves a slight energy saving over \textit{Linux-BayOp}. This result suggests an interesting compromise between enabling a degree of energy savings that controlling ITR-delay provides to a computationally-driven network application versus abandoning ITR-delay control so that Bayesian optimization can focus on tuning DVFS to maximize energy savings.

%% file: tailbench.tex
\subsection{Black-Box Generality: Diverse Apps and Hardware}
\label{sec:tailbench}
\begin{table}[t]
\small
\begin{tabular}{|c|c|c|c|c|}
    \hline
  Name & CPU & Cores & NIC & RAM\\ \hline
     N0 & Intel E5-2640 & 8 & Mellanox 25GbE & 62GB\\ \hline
     N1 & Intel E5-2660 & 20 & Solarflare 10GbE & 128GB\\ \hline
     N2 & AMD EPYC 7452 & 32 & Mellanox 40GbE & 128GB\\ \hline
     N3 & Ampere ARMv8 & 80 & Mellanox 25GbE & 124 GB \\ \hline
\end{tabular}
\caption{\small Different hardware explored to run Tailbench.}
\label{table:hwsw}
\vspace{-0.2in}
\end{table}

In this section, we further demonstrate the versatility of the controller by applying it to optimize energy efficiency for three applications from Tailbench~\cite{tailbench}. Our motivation was to reveal how externally controlling interrupt coalescing and CPU frequency can be applied agnostically on hardware even across multi-generational divides\footnote{Scripts to reproduce results at https://anonymous.4open.science/r/bayop-188B}.

\subsubsection{Experimental Setup}
For these experiments, we selected four hardware platforms as shown in \cref{table:hwsw}. Nodes N0, N1, and N2 are provided by CloudLab~\cite{cloudlab} and we disable hyperthreads and TurboBoost on all processors to minimize system noise. For each node type, we create a cluster consisting of a single server node, three client nodes that generate traffic to the server node, and an external bootstrap node that launches experiments and runs the \texttt{BayOp} controller to tune interrupt coalescing (ITR-delay) and CPU frequency (DVFS) on the server. All of the nodes were running Linux 5.15 kernel; we only examined Linux as EbbRT does not have the necessary device driver support for Solarflare and Mellanox NICs. Notably, while we used \texttt{ethtool} to set static interrupt rates across all three NICs in this paper, the fundamental implementation may be different depending on the hardware's capability. On the Intel processors, we use the RAPL hardware registers~\cite{rapl} to report its dynamic energy use while for AMD, we use \texttt{amd\_energy} hardware monitor driver~\cite{amdenergy}.  

Node N3 is another experimental node that runs Linux 6.4.13 but we could only get a single client node to generate traffic\footnote{Due to the computation-heavy nature of Tailbench applications, we found this was still able to saturate the single server}. The ARMv8 server provided \texttt{xgene-hwmon}~\cite{armxgene} tool that enabled us to report its power readings. 

For each hardware category, we selected applications from Tailbench~\cite{tailbench}, each designed to fulfill distinct SLA objectives. These applications encompassed \textbf{img-dnn}, a handwriting recognition program built on OpenCV; \textbf{sphinx}, an open-source search engine; and \textbf{xapian}, a speech recognition system. These applications both represent a diverse suite of benchmarks in contrast to the previous examples from our study as well as providing new SLA objectives in the order of milliseconds to seconds. Overall, these selections allowed us to assess the impact of different SLAs and hardware platforms.

\subsubsection{Experimental Results}
In our experiments, we observed that the controller consistently achieves energy savings ranging from 5\% to 36\%, depending on the specific combination of software and hardware. Importantly, our findings underscore the fundamental nature of these two mechanisms, which can be effectively applied across a variety of hardware platforms in different SLA-driven application domains. Further, we found that the generic architecture of our controller meant that it was straightforward to simply deploy this technique in new hardware environments as long as it provided support for energy readings and exposed control of interrupt coalescing and CPU frequency. 

Fig.~\ref{fig:xl170} depicts the resulting energy consumption for Tailbench; we normalize the energy usage relative to the default Linux configurations under different scenarios:
\begin{enumerate}
    \item We selected representative offered loads of 40\% and 80\% of each hardware platform's peak QPS capacity for running the respective applications.
    \item For each of these offered loads, we applied two distinct SLA objectives tailored to each application, as indicated by the labels in each figure. These SLA objectives were derived from default values provided by the authors of Tailbench~\cite{tailbench}.
\end{enumerate}

However, it is worth pointing out that the controller's ability to adapt to applications and offered loads is heavily influenced by the hardware's ability to offer a range of configurations for exploration within this space. One can see an example of this for \textbf{APP: img-dnn} in \textbf{Node: N2} where it did not manage to find an ITR-delay, DVFS pair that managed to further reduce energy consumption. We hypothesize this stems from a combination of the application type as well as the DVFS settings provided by the AMD EPYC 7452 processor. The processor uses AMD's Collaborative Processor Performance Control (CPPC) interface~\cite{amdpstate}, which is an abstracted performance value that isn't tied to specific a CPU frequency; further, we were limited to only three settings in contrast to the hundreds and thousands available on the other processors. However, this limitation can also be mitigated by newer processors that support the AMD P-state EPP~\cite{amdepp} driver, providing finer-grained CPU frequency settings.

\begin{figure}[!htb]
\centering
    \includegraphics[width=0.45\textwidth]{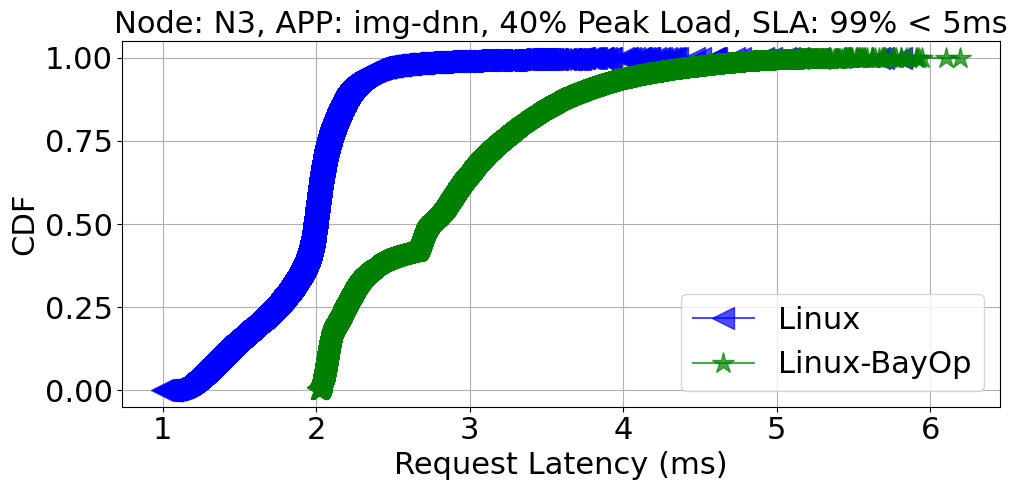}
    \caption{\small CDF of per request latency between Linux and Linux-BayOp from a single Tailbench application.}
    \label{fig:cdf1}
    \vspace{-0.2in}
\end{figure}

To delve into the energy gains 
we detail the CDF of an example Tailbench application in \cref{fig:cdf1}. In this figure, we illustrate the per-request latency as provided by Tailbench when running the img-dnn application on our ARMv8 server (note this is at a particular peak load and SLA). As the figure shows, the overall request latency of \textit{Linux-BayOP} is about 2X worse than Linux as the controller chose energy-efficient ITR-delay and DVFS settings. While we found Linux was able to support this workload with a 99\% latency of 2.8ms, \textit{Linux-BayOp} was still able to meet the SLA at 99\% latency of 4.8ms while saving 31\% energy.

%% file: related.tex
\section{Related Work}
\label{sec:related}
Our work falls within a wider space of research on energy proportional computation in datacenters~\cite{energyproportion, warehouse-power, 268014}. Much of this research stems from the challenges of improving the performance of network-bound data center workloads~\cite{large-scale-mapreduce, mica, zygos} while keeping energy consumption at bay. These challenges can be attributed to complex diurnal trends that are characteristic of datacenter-level utilization, whereby idle time is common and must be optimized for~\cite{hotpower2008, powernap, napsac} while simultaneously maintaining the ability to support high-utilization peaks and strict latency constraints ~\cite{Dynamo, SmoothOperator, oldi-pegasus, adrenaline, rubik, eurosys14, zygos, peafowl, 7425206, 10.1145/2830772.2830779, dreamweaver, dynsleep, udpm}. Our goal was to gain better insight into these impacts on application performance and energy when ITR-delay and DVFS settings are precisely controlled. While prior work has shown how SLA headrooms can be exploited to minimize the overall energy consumption of a system~\cite{Dynamo, SmoothOperator, oldi-pegasus, adrenaline, rubik, eurosys14, zygos, peafowl, 7425206, 10.1145/2830772.2830779, dreamweaver, dynsleep, udpm}, our controller demonstrates this process can be automated and customized on a wider range of hardware and applications than previously shown.

There is a wide range of work that targets energy proportionality with a focus on designing OS policies and mechanisms for power management. Most of this work presents hardware level optimizations that manipulate processor speed mechanisms such as DVFS ~\cite{10.5555/2523721.2523732,10.1145/381677.381702,cpufreq_governor,4273098,packandcap,10.1109/MICRO.2006.8,1598114,10.1145/1629911.1629926,4658633,4343825,10.1109/IGCC.2011.6008552,10.1145/1241601.1241609, slowdownorsleep,4228267, mootaz}, processor power limiting mechanisms such as RAPL~\cite{intel_rapl, heracles, SmoothOperator,oldi-pegasus, Dynamo,PerAppPower,powercap}, and idle power states~\cite{cpuidle_policy,peafowl, udpm,6983037,dreamweaver, pacingtoidle} (c-states) by applying feedback control mechanisms and relying on activity models. The authors of ~\cite{heracles} and ~\cite{PerAppPower} go a step further, exploring and characterizing the interference of co-located latency-critical versus best-effort tasks and high versus low CPU demand tasks when subject to energy tuning via DVFS and RAPL. In doing so, they highlight limitations in using hardware features alone for power management. Our work builds on this observation and asserts that specialization in the OS stack also plays a critical role in attaining even more energy efficiency. 

Modern hardware components and software stacks expose a large number of parameters that govern internal system operations and interactions. There is a lot of work on defining heuristics to control hardware parameters that impact performance and energy consumption ~\cite{10.1145/2626401.2626411, 7108419, caloree, 10.1145/2775054.2694373, arxiv.2112.07010, heuristics_0}. In recent years, there has been an explosion in using ML-based techniques \cite{Wu2023, 10.1145/3307650.3326633} to uncover more subtle system heuristics for resource management ~\cite{10.5555/1855591.1855592, 4771801, 10.1145/1995896.1995927, paragon, quasar, 5695560, 10.1145/2815400.2815403, 10.1145/1394608.1382172, caloree, 10.1145/2775054.2694373, carat, flicker, koala, 6522303}, hardware and system configuration ~\cite{Siblingrivalry, 7207248, mem_cocktail, 5695560, 10.5555/1855591.1855592, 10.1145/1168857.1168881, 1598114, 10.1145/2829950, 10.1145/3035918.3064029, 6493638, 6730744, 7551432, bestconfig}, high-performance computing ~\cite{10.1145/1229428.1229479, 10.1145/2775054.2694373, 10.1145/1394608.1382172, 10.1145/2872362.2872375, Siblingrivalry, 1598114, 10.1145/3453483.3454109}, and data-center-scale applications ~\cite{paragon, quasar, 6730744, 7207248, 10.1145/3035918.3064029, bestconfig, 4061117, nurd, 10.1145/3468264.3468603, 10.1145/3342195.3387520, 10.1145/3173162.3173206, cello}. Though ML is a natural solution for domains like image, video, and audio processing, the complexity of computer systems often requires extensive expertise to map systems problems to ML tasks. Therefore, prior research has either been limited to simulators~\cite{4771801, 10.1145/1995896.1995927, 5695560, 10.1145/1394608.1382172, 1635956, mem_cocktail, 1598114, 10.1145/1839667.1839670, 6493638, 7851506, nurd} or focused only on software parameters only~\cite{paragon, quasar, carat, 6522303, 6730744, Siblingrivalry, 10.1145/2872362.2872375, bestconfig, cello, 10.1145/3173162.3173206, 10.1145/3342195.3387520}. Instead, our work is the first to apply an ML technique towards exploiting stability in offered loads to find energy-efficient "sweet spots". Our work on finding settings for ITR-delay and DVFS for SLA-driven network applications is most similar to Co-PI~\cite{9248059}. Their approach focuses on the hardware and software specific nature of optimizing ITR-Delay and DVFS on an Intel platform; through off-line profiling, they construct lookup tables indexed base on three coarse gain load categories (low, medium and high). Our work demonstrates how external control of interrupt coalescing and CPU frequency are fundamental mechanisms that can be generally applied across offered load, application, OS, and hardware. Further, we demonstrate how Bayesian optimization can be used to dynamically reduce energy use across a variety of SLA objectives on a live server.

%% file: conclusion.tex
\section{Conclusion}
\label{sec:conc}
Our work seeks to validate a set of conjectures about how combining queuing and processing efficiency can result in diverse set of energy-efficient system settings. Further, to confirm our conjecture that one can exploit stable demand curves in SLA-driven applications; we have also proposed an example controller design that utilizes a black-box search strategy to automatically find these "sweet spots". Our results demonstrate its applicability across offered loads, applications, OSes and even hardware.

%% file: appendix.tex
\section{Appendix}
\label{sec:appendix}

\begin{figure}[!htb]
\centering
\begin{subfigure}{.45\textwidth}
  \includegraphics[width=\linewidth]{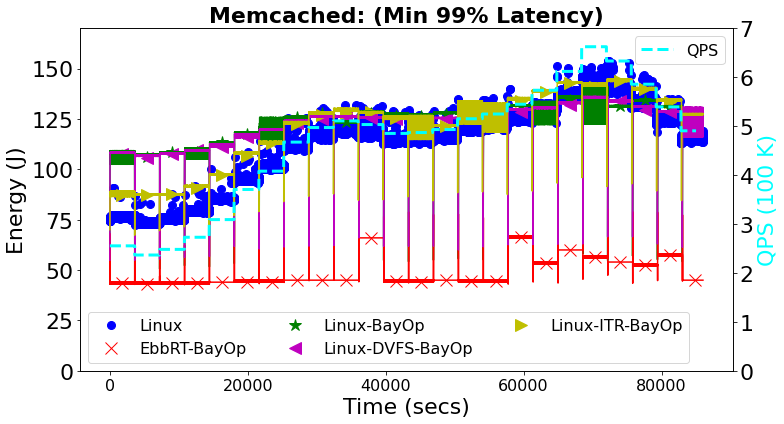}
\end{subfigure}
\begin{subfigure}{.45\textwidth}
  \includegraphics[width=\linewidth]{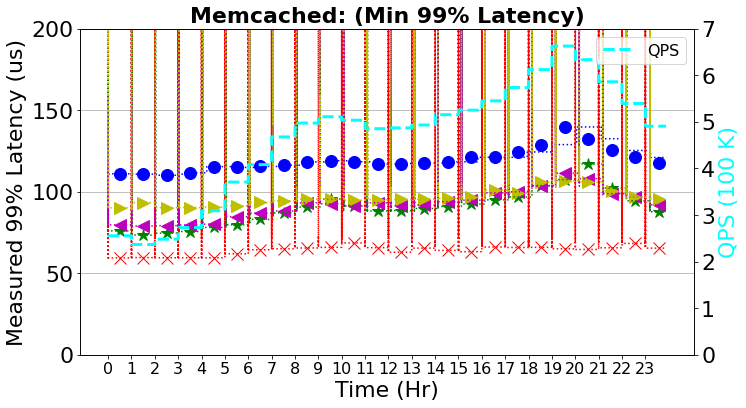}
\end{subfigure}
\caption{\small Controller applied to optimize \textbf{only} for minimizing 99\% tail latency for memcached. We show show the energy per second consumption on the left figure and the measured latency on the right.}
\label{fig:bayesopt_mcd_min99}
\end{figure}

\subsection{Optimizing for Latency in Memcached}

In \cref{fig:bayesopt_mcd_min99} we demonstrate the flexibility of the controller's optimization criteria through a change of its reward penalty function to focus on minimizing tail latency instead at a cost to greater energy use. In this case, the function is simplified to $Rp = m\_latency$ in order to reflect performance optimization instead. Fig.~\ref{fig:bayesopt_mcd_min99} illustrates that the Bayesian process was also able to consistently performance-focused ITR-delay,  DVFS settings that lowered the 99\% tail latency by up to 30\% in Linux, however at a higher energy cost of up to 40\%. This result also demonstrates the potential of exploring alternate reward functions that may consist of different combinations of performance and energy criteria.

\begin{figure*}[!htb]
\centering
\begin{subfigure}{.45\textwidth}
  \includegraphics[width=\linewidth]{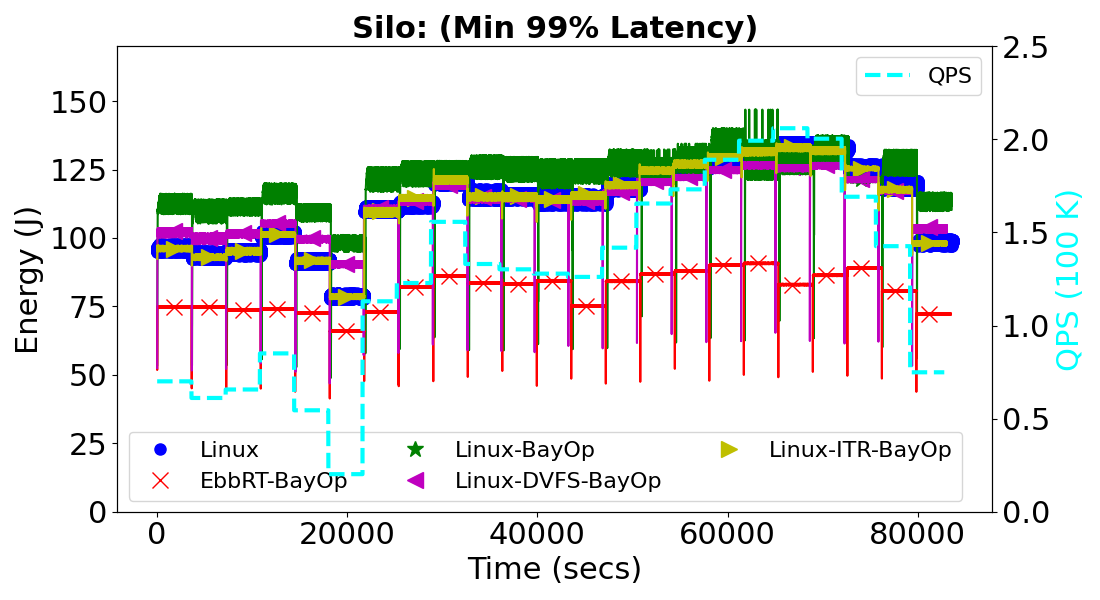}
\end{subfigure}%
\begin{subfigure}{.45\textwidth}
  \includegraphics[width=\linewidth]{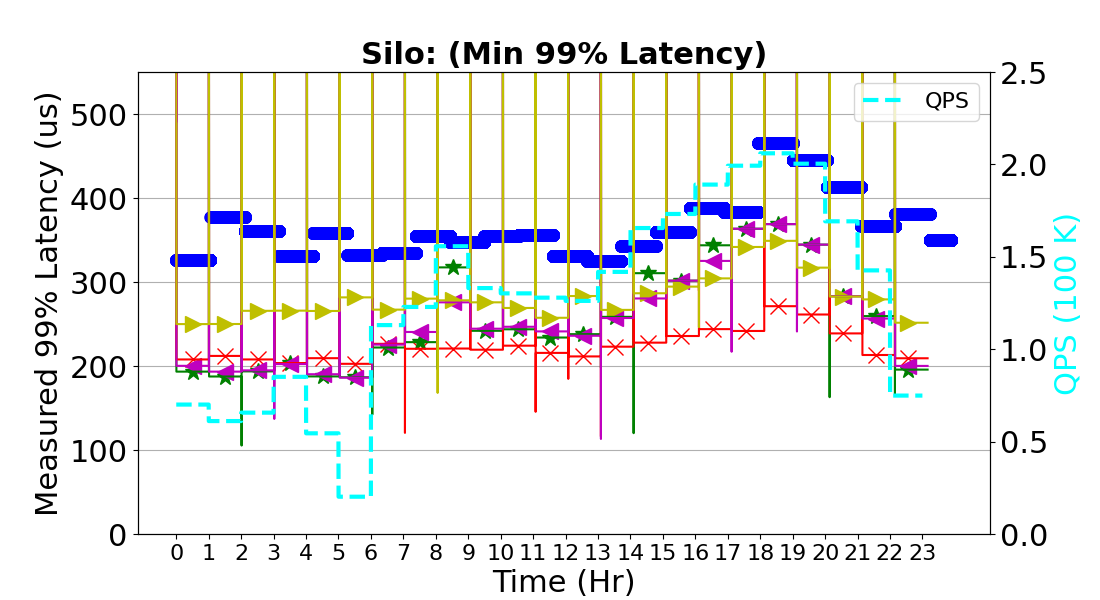}
\end{subfigure}
\caption{\small Controller applied to optimize \textbf{only} for minimizing 99\% tail latency in Silo. We show show the energy per second consumption on the left figure and the measured latency on the right.
}
\label{fig:bayesopt_mcdsilo_min99}
\end{figure*}

\subsection{Optimizing for Latency in Silo} Fig.~\ref{fig:bayesopt_mcdsilo_min99} shows that similar to memcached, the controller can still largely lower overall 99\% latency by 50 \% while increasing its overall energy use for a new application across both OSes.

%% file: main.bbl
\begin{thebibliography}{100}

\bibitem{intelethtool}
{Improving Measured Latency in Linux for Intel® 82575/82576 or
  X540/82598/82599 Ethernet Controllers}.
\newblock
  \url{https://www.intel.com/content/www/us/en/embedded/products/networking/82575-82576-82598-82599-ethernet\\-controllers-latency-appl-note.html}.

\bibitem{armxgene}
{Kernel driver xgene-hwmon}.
\newblock \url{https://docs.kernel.org/hwmon/xgene-hwmon.html}.

\bibitem{Siblingrivalry}
Jason Ansel, Maciej Pacula, Yee~Lok Wong, Cy~Chan, Marek Olszewski, Una-May
  O'Reilly, and Saman Amarasinghe.
\newblock Siblingrivalry: Online autotuning through local competitions.
\newblock In {\em Proceedings of the 2012 International Conference on
  Compilers, Architectures and Synthesis for Embedded Systems}, CASES '12, page
  91–100, New York, NY, USA, 2012. Association for Computing Machinery.

\bibitem{rumpkernel}
{Antti Kantee, Justin Cormack}.
\newblock {Rump Kernels: No OS? No Problem!}
\newblock
  \url{https://www.usenix.org/publications/login/october-2014-vol-39-no-5}.

\bibitem{armdvfs}
{ARM}.
\newblock
  \url{https://developer.arm.com/documentation/den0013/d/Power-Management}.

\bibitem{peafowl}
Esmail Asyabi, Azer Bestavros, Erfan Sharafzadeh, and Timothy Zhu.
\newblock Peafowl: In-application cpu scheduling to reduce power consumption of
  in-memory key-value stores.
\newblock In {\em Proceedings of the 11th ACM Symposium on Cloud Computing},
  SoCC '20, page 150–164, New York, NY, USA, 2020. Association for Computing
  Machinery.

\bibitem{workloadanalysisfacebook}
{Atikoglu, Berk and Xu, Yuehai and Frachtenberg, Eitan and Jiang, Song and
  Paleczny, Mike}.
\newblock {Workload Analysis of a Large-scale Key-value Store}.
\newblock In {\em {Proceedings of the 12th ACM SIGMETRICS/PERFORMANCE Joint
  International Conference on Measurement and Modeling of Computer Systems}},
  SIGMETRICS '12, pages 53--64, New York, NY, USA, 2012. ACM.

\bibitem{ax}
{Ax:Adaptive Experimentation Platform}.
\newblock \url{https://ax.dev/}.

\bibitem{Bakshy2018AEAD}
Eytan Bakshy, Lili Dworkin, Brian Karrer, Konstantin Kashin, Benjamin Letham,
  Ashwin Murthy, and Shaun Singh.
\newblock Ae: A domain-agnostic platform for adaptive experimentation.
\newblock 2018.

\bibitem{WebSearch}
Luiz~Andre Barroso, Jeffrey Dean, and Urs Hölzle.
\newblock Web search for a planet: The google cluster architecture.
\newblock {\em IEEE Micro}, 23:22--28, 2003.

\bibitem{Barroso:2009:DCI:1643608}
Luiz~Andre Barroso and Urs Hoelzle.
\newblock {\em The Datacenter As a Computer: An Introduction to the Design of
  Warehouse-Scale Machines}.
\newblock Morgan and Claypool Publishers, 1st edition, 2009.

\bibitem{energyproportion}
{Barroso, Luiz Andr\'{e} and H\"{o}lzle, Urs}.
\newblock The case for energy-proportional computing.
\newblock {\em Computer}, 40(12):33–37, December 2007.

\bibitem{ix}
Adam Belay, George Prekas, Ana Klimovic, Samuel Grossman, Christos Kozyrakis,
  and Edouard Bugnion.
\newblock Ix: A protected dataplane operating system for high throughput and
  low latency.
\newblock In {\em Proceedings of the 11th USENIX Conference on Operating
  Systems Design and Implementation}, OSDI'14, page 49–65, USA, 2014. USENIX
  Association.

\bibitem{4771801}
Ramazan Bitirgen, Engin Ipek, and Jose~F. Martinez.
\newblock Coordinated management of multiple interacting resources in chip
  multiprocessors: A machine learning approach.
\newblock In {\em 2008 41st IEEE/ACM International Symposium on
  Microarchitecture}, pages 318--329, 2008.

\bibitem{10.1109/40.888701}
David~M. Brooks, Pradip Bose, Stanley~E. Schuster, Hans Jacobson, Prabhakar~N.
  Kudva, Alper Buyuktosunoglu, John-David Wellman, Victor Zyuban, Manish Gupta,
  and Peter~W. Cook.
\newblock Power-aware microarchitecture: Design and modeling challenges for
  next-generation microprocessors.
\newblock {\em IEEE Micro}, 20(6):26–44, November 2000.

\bibitem{10.1145/2626401.2626411}
Aaron Carroll and Gernot Heiser.
\newblock Mobile multicores: Use them or waste them.
\newblock {\em SIGOPS Oper. Syst. Rev.}, 48(1):44–48, may 2014.

\bibitem{7207248}
Chi-Ou Chen, Ye-Qi Zhuo, Chao-Chun Yeh, Che-Min Lin, and Shih-Wei Liao.
\newblock Machine learning-based configuration parameter tuning on hadoop
  system.
\newblock In {\em 2015 IEEE International Congress on Big Data}, pages
  386--392, 2015.

\bibitem{10.1145/1995896.1995927}
Jian Chen and Lizy~Kurian John.
\newblock Predictive coordination of multiple on-chip resources for chip
  multiprocessors.
\newblock In {\em Proceedings of the International Conference on
  Supercomputing}, ICS '11, page 192–201, New York, NY, USA, 2011.
  Association for Computing Machinery.

\bibitem{large-scale-mapreduce}
Yanpei Chen, Sara Alspaugh, Dhruba Borthakur, and Randy Katz.
\newblock Energy efficiency for large-scale mapreduce workloads with
  significant interactive analysis.
\newblock In {\em Proceedings of the 7th ACM European Conference on Computer
  Systems}, EuroSys ’12, page 43–56, New York, NY, USA, 2012. Association
  for Computing Machinery.

\bibitem{1635956}
Seungryul Choi and D.~Yeung.
\newblock Learning-based smt processor resource distribution via hill-climbing.
\newblock In {\em 33rd International Symposium on Computer Architecture
  (ISCA'06)}, pages 239--251, 2006.

\bibitem{udpm}
C.~Chou, L.~N. Bhuyan, and D.~Wong.
\newblock Î¼dpm: Dynamic power management for the microsecond era.
\newblock In {\em 2019 IEEE International Symposium on High Performance
  Computer Architecture (HPCA)}, pages 120--132, Los Alamitos, CA, USA, feb
  2019. IEEE Computer Society.

\bibitem{dynsleep}
Chih-Hsun Chou, Daniel Wong, and Laxmi~N. Bhuyan.
\newblock Dynsleep: Fine-grained power management for a latency-critical data
  center application.
\newblock In {\em Proceedings of the 2016 International Symposium on Low Power
  Electronics and Design}, ISLPED '16, page 212–217, New York, NY, USA, 2016.
  Association for Computing Machinery.

\bibitem{packandcap}
Ryan Cochran, Can Hankendi, Ayse~K. Coskun, and Sherief Reda.
\newblock {Pack \& Cap: Adaptive DVFS and Thread Packing under Power Caps}.
\newblock In {\em Proceedings of the 44th Annual IEEE/ACM International
  Symposium on Microarchitecture}, MICRO-44, page 175–185, New York, NY, USA,
  2011. Association for Computing Machinery.

\bibitem{netflixmcd}
{Daniel Ellis}.
\newblock
  \url{https://netflixtechblog.com/ephemeral-volatile-caching-\\in-the-cloud-8eba7b124589}.

\bibitem{redditmcd}
{Daniel Ellis}.
\newblock
  \url{https://web.archive.org/web/20210205121832/https://redditblog.com/2017/1/17/caching-at-reddit/}.

\bibitem{rapl}
Howard David, Eugene Gorbatov, Ulf~R. Hanebutte, Rahul Khanna, and Christian
  Le.
\newblock Rapl: Memory power estimation and capping.
\newblock In {\em Proceedings of the 16th ACM/IEEE International Symposium on
  Low Power Electronics and Design}, ISLPED ’10, page 189–194, New York,
  NY, USA, 2010. Association for Computing Machinery.

\bibitem{paragon}
Christina Delimitrou and Christos Kozyrakis.
\newblock Paragon: Qos-aware scheduling for heterogeneous datacenters.
\newblock In {\em Proceedings of the Eighteenth International Conference on
  Architectural Support for Programming Languages and Operating Systems},
  ASPLOS '13, page 77–88, New York, NY, USA, 2013. Association for Computing
  Machinery.

\bibitem{quasar}
Christina Delimitrou and Christos Kozyrakis.
\newblock Quasar: Resource-efficient and qos-aware cluster management.
\newblock In {\em Proceedings of the 19th International Conference on
  Architectural Support for Programming Languages and Operating Systems},
  ASPLOS '14, page 127–144, New York, NY, USA, 2014. Association for
  Computing Machinery.

\bibitem{mem_cocktail}
Zhaoxia Deng, Lunkai Zhang, Nikita Mishra, Henry Hoffmann, and Frederic~T.
  Chong.
\newblock Memory cocktail therapy: A general learning-based framework to
  optimize dynamic tradeoffs in nvms.
\newblock In {\em Proceedings of the 50th Annual IEEE/ACM International
  Symposium on Microarchitecture}, MICRO-50 '17, page 232–244, New York, NY,
  USA, 2017. Association for Computing Machinery.

\bibitem{10.1145/3307650.3326633}
Yi~Ding, Nikita Mishra, and Henry Hoffmann.
\newblock Generative and multi-phase learning for computer systems
  optimization.
\newblock In {\em Proceedings of the 46th International Symposium on Computer
  Architecture}, ISCA '19, page 39–52, New York, NY, USA, 2019. Association
  for Computing Machinery.

\bibitem{10.1145/3468264.3468603}
Yi~Ding, Ahsan Pervaiz, Michael Carbin, and Henry Hoffmann.
\newblock Generalizable and interpretable learning for configuration
  extrapolation.
\newblock In {\em Proceedings of the 29th ACM Joint Meeting on European
  Software Engineering Conference and Symposium on the Foundations of Software
  Engineering}, ESEC/FSE 2021, page 728–740, New York, NY, USA, 2021.
  Association for Computing Machinery.

\bibitem{nurd}
Yi~Ding, Avinash Rao, Hyebin Song, Rebecca Willett, and Henry Hoffmann.
\newblock Nurd: Negative-unlabeled learning for online datacenter straggler
  prediction, 2022.

\bibitem{cello}
Yi~Ding, Alex Renda, Ahsan Pervaiz, Michael Carbin, and Henry Hoffmann.
\newblock Cello: Efficient computer systems optimization with predictive early
  termination and censored regression, 2022.

\bibitem{cpufreq_governor}
{Dominik Brodowski, Nico Golde, Rafael J. Wysocki, Viresh Kumar}.
\newblock {CPU frequency and voltage scaling code in the Linux(TM) kernel}.
\newblock
  \url{https://www.kernel.org/doc/Documentation/cpu-freq/governors.txt}.

\bibitem{arxiv.2112.07010}
Han Dong, Sanjay Arora, Yara Awad, Tommy Unger, Orran Krieger, and Jonathan
  Appavoo.
\newblock Slowing down for performance and energy: An os-centric study in
  network driven workloads. https://arxiv.org/abs/2112.07010, 2021.

\bibitem{5695560}
Christophe Dubach, Timothy~M. Jones, Edwin~V. Bonilla, and Michael~F.P.
  O'Boyle.
\newblock A predictive model for dynamic microarchitectural adaptivity control.
\newblock In {\em 2010 43rd Annual IEEE/ACM International Symposium on
  Microarchitecture}, pages 485--496, 2010.

\bibitem{cloudlab}
Dmitry Duplyakin, Robert Ricci, Aleksander Maricq, Gary Wong, Jonathon Duerig,
  Eric Eide, Leigh Stoller, Mike Hibler, David Johnson, Kirk Webb, Aditya
  Akella, Kuangching Wang, Glenn Ricart, Larry Landweber, Chip Elliott, Michael
  Zink, Emmanuel Cecchet, Snigdhaswin Kar, and Prabodh Mishra.
\newblock The design and operation of {CloudLab}.
\newblock In {\em Proceedings of the {USENIX} Annual Technical Conference
  (ATC)}, pages 1--14, July 2019.

\bibitem{mootaz}
Mootaz Elnozahy, Michael Kistler, and Ramakrishnan Rajamony.
\newblock Energy conservation policies for web servers.
\newblock In {\em Proceedings of the 4th Conference on USENIX Symposium on
  Internet Technologies and Systems - Volume 4}, USITS'03, page~8, USA, 2003.
  USENIX Association.

\bibitem{warehouse-power}
Xiaobo Fan, Wolf-Dietrich Weber, and Luiz~Andre Barroso.
\newblock Power provisioning for a warehouse-sized computer.
\newblock In {\em Proceedings of the 34th Annual International Symposium on
  Computer Architecture}, ISCA ’07, page 13–23, New York, NY, USA, 2007.
  Association for Computing Machinery.

\bibitem{memcached}
Brad Fitzpatrick.
\newblock {Distributed Caching with Memcached}.
\newblock {\em Linux Journal}, 2004(124):5, August 2004.

\bibitem{10.1145/381677.381702}
Kriszti\'{a}n Flautner, Steve Reinhardt, and Trevor Mudge.
\newblock Automatic performance setting for dynamic voltage scaling.
\newblock In {\em Proceedings of the 7th Annual International Conference on
  Mobile Computing and Networking}, MobiCom '01, page 260–271, New York, NY,
  USA, 2001. Association for Computing Machinery.

\bibitem{frazier}
Peter~I. Frazier.
\newblock A tutorial on bayesian optimization, 2018.

\bibitem{4228267}
Vincent~W. Freeh, Tyler~K. Bletsch, and Freeman~L. Rawson.
\newblock Scaling and packing on a chip multiprocessor.
\newblock In {\em 2007 IEEE International Parallel and Distributed Processing
  Symposium}, pages 1--8, 2007.

\bibitem{10.5555/1855591.1855592}
Archana Ganapathi, Kaushik Datta, Armando Fox, and David Patterson.
\newblock A case for machine learning to optimize multicore performance.
\newblock In {\em Proceedings of the First USENIX Conference on Hot Topics in
  Parallelism}, HotPar'09, page~1, USA, 2009. USENIX Association.

\bibitem{garnett_bayesoptbook_2022}
Roman Garnett.
\newblock {\em {Bayesian Optimization}}.
\newblock Cambridge University Press, 2022.
\newblock in preparation.

\bibitem{4343825}
Rong Ge, Xizhou Feng, Wu-chun Feng, and Kirk~W. Cameron.
\newblock Cpu miser: A performance-directed, run-time system for power-aware
  clusters.
\newblock In {\em 2007 International Conference on Parallel Processing (ICPP
  2007)}, pages 18--18, 2007.

\bibitem{PerAppPower}
Akhil Guliani and Michael~M. Swift.
\newblock Per-application power delivery.
\newblock In {\em Proceedings of the Fourteenth EuroSys Conference 2019},
  EuroSys '19, New York, NY, USA, 2019. Association for Computing Machinery.

\bibitem{gupta2020chasing}
Udit Gupta, Young~Geun Kim, Sylvia Lee, Jordan Tse, Hsien-Hsin~S. Lee, Gu-Yeon
  Wei, David Brooks, and Carole-Jean Wu.
\newblock Chasing carbon: The elusive environmental footprint of computing,
  2020.

\bibitem{hanappliance}
{Han Dong, Jonathan Appavoo}.
\newblock {A Tutorial on Building Custom Linux Appliances}.
\newblock
  \url{https://www.usenix.org/publications/loginonline/building-linux-appliances}.

\bibitem{10.1145/2815400.2815403}
Henry Hoffmann.
\newblock Jouleguard: Energy guarantees for approximate applications.
\newblock In {\em Proceedings of the 25th Symposium on Operating Systems
  Principles}, SOSP '15, page 198–214, New York, NY, USA, 2015. Association
  for Computing Machinery.

\bibitem{573184}
M.~Horowitz, T.~Indermaur, and R.~Gonzalez.
\newblock Low-power digital design.
\newblock In {\em Proceedings of 1994 IEEE Symposium on Low Power Electronics},
  pages 8--11, 1994.

\bibitem{adrenaline}
C.~{Hsu}, Y.~{Zhang}, M.~A. {Laurenzano}, D.~{Meisner}, T.~{Wenisch},
  J.~{Mars}, L.~{Tang}, and R.~G. {Dreslinski}.
\newblock Adrenaline: Pinpointing and reining in tail queries with quick
  voltage boosting.
\newblock In {\em 2015 IEEE 21st International Symposium on High Performance
  Computer Architecture (HPCA)}, pages 271--282, 2015.

\bibitem{SmoothOperator}
Chang-Hong Hsu, Qingyuan Deng, Jason Mars, and Lingjia Tang.
\newblock Smoothoperator: Reducing power fragmentation and improving power
  utilization in large-scale datacenters.
\newblock In {\em Proceedings of the Twenty-Third International Conference on
  Architectural Support for Programming Languages and Operating Systems},
  ASPLOS '18, pages 535--548, New York, NY, USA, 2018. ACM.

\bibitem{amdpstate}
{Huang Rui}.
\newblock {amd-pstate CPU Performance Scaling Driver}.
\newblock \url{https://docs.kernel.org/admin-guide/pm/amd-pstate.html}.

\bibitem{7108419}
C.~Imes, D.~K. Kim, M.~Maggio, and H.~Hoffmann.
\newblock Poet: a portable approach to minimizing energy under soft real-time
  constraints.
\newblock In {\em 2015 IEEE Real-Time and Embedded Technology and Applications
  Symposium (RTAS)}, pages 75--86, Los Alamitos, CA, USA, apr 2015. IEEE
  Computer Society.

\bibitem{intel_manual}
Intel.
\newblock {Intel® 64 and IA-32 Architectures Software Developer’s Manual
  Volume}.
\newblock
  \url{https://www.intel.com/content/dam/www/public/us/en/documents/manuals/}.

\bibitem{intel_rapl}
Intel.
\newblock {Intel® 64 and IA-32 Architectures Software Developer’s Manual
  Volume 3B:System Programming Guide, Part 2}.
\newblock
  \url{https://www.intel.com/content/dam/www/public/us/en/documents/manuals/64-ia-32-architectures-software-developer-\\vol-3b-part-2-manual.pdf}.

\bibitem{intel_msr}
Intel.
\newblock {Intel® 64 and IA-32 Architectures Software Developer’s Manual
  Volume 3C:System Programming Guide, Part 3}.
\newblock
  \url{https://www.intel.com/content/dam/www/public/us/en/documents/manuals/64-ia-32-architectures-software-developer-\\vol-3c-part-3-manual.pdf}.

\bibitem{intelitr}
{Intel}.
\newblock {Tuning Throughput Performance for Intel® Ethernet Adapters}.
\newblock
  \url{https://www.intel.com/content/www/us/en/support/articles/000005811/network-and-i-o/ethernet-products.html}.

\bibitem{82599}
{Intel 82599 10 Gigabit Ethernet Controller: Datasheet}.
\newblock
  \url{https://www.intel.com/content/www/us/en/embedded/products/networking/82599-10-gbe-controller-datasheet.html}.

\bibitem{10.1145/1394608.1382172}
Engin Ipek, Onur Mutlu, Jos\'{e}~F. Mart\'{\i}nez, and Rich Caruana.
\newblock Self-optimizing memory controllers: A reinforcement learning
  approach.
\newblock {\em SIGARCH Comput. Archit. News}, 36(3):39–50, jun 2008.

\bibitem{10.1109/MICRO.2006.8}
Canturk Isci, Alper Buyuktosunoglu, Chen-Yong Cher, Pradip Bose, and Margaret
  Martonosi.
\newblock An analysis of efficient multi-core global power management policies:
  Maximizing performance for a given power budget.
\newblock In {\em Proceedings of the 39th Annual IEEE/ACM International
  Symposium on Microarchitecture}, MICRO 39, page 347–358, USA, 2006. IEEE
  Computer Society.

\bibitem{mutilate}
{J. Leverich}.
\newblock {Mutilate: high performance memcached load generator}.
\newblock \url{https://github.com/leverich/mutilate}.

\bibitem{mtcp}
EunYoung Jeong, Shinae Wood, Muhammad Jamshed, Haewon Jeong, Sunghwan Ihm,
  Dongsu Han, and KyoungSoo Park.
\newblock mtcp: a highly scalable user-level {TCP} stack for multicore systems.
\newblock In {\em 11th {USENIX} Symposium on Networked Systems Design and
  Implementation ({NSDI} 14)}, pages 489--502, Seattle, WA, April 2014.
  {USENIX} Association.

\bibitem{cacheWorkload-OSDI20}
Rashmi~Vinayak Juncheng~Yang, Yao~Yue.
\newblock A large scale analysis of hundreds of in-memory cache clusters at
  twitter.
\newblock In {\em 14th USENIX Symposium on Operating Systems Design and
  Implementation (OSDI 20)}. {USENIX} Association, November 2020.

\bibitem{intelinterruptmoderation}
{Kan Liang, Andi Kleen, and Jesse Brandenburg}.
\newblock {Improve Network Performance By Setting Per-queue Interrupt
  Moderation In Linux}.
\newblock \url{https://01.org/linux-interrupt-moderation}.

\bibitem{6983037}
Svilen Kanev, Kim Hazelwood, Gu-Yeon Wei, and David Brooks.
\newblock Tradeoffs between power management and tail latency in
  warehouse-scale applications.
\newblock In {\em 2014 IEEE International Symposium on Workload
  Characterization (IISWC)}, pages 31--40, 2014.

\bibitem{9248059}
Ki-Dong Kang, Hyungwon Park, Gyeongseo Park, and Daehoon Kim.
\newblock Co-adjusting voltage/frequency state and interrupt rate for improving
  energy-efficiency of latency-critical applications.
\newblock {\em IEEE Access}, 8:201028--201039, 2020.

\bibitem{rubik}
H.~{Kasture}, D.~B. {Bartolini}, N.~{Beckmann}, and D.~{Sanchez}.
\newblock Rubik: Fast analytical power management for latency-critical systems.
\newblock In {\em 2015 48th Annual IEEE/ACM International Symposium on
  Microarchitecture (MICRO)}, pages 598--610, 2015.

\bibitem{tailbench}
Harshad Kasture and Daniel Sanchez.
\newblock Tailbench: a benchmark suite and evaluation methodology for
  latency-critical applications.
\newblock In {\em 2016 IEEE International Symposium on Workload
  Characterization (IISWC)}, pages 1--10, 2016.

\bibitem{flexnic}
Antoine Kaufmann, SImon Peter, Naveen~Kr. Sharma, Thomas Anderson, and Arvind
  Krishnamurthy.
\newblock High performance packet processing with flexnic.
\newblock In {\em Proceedings of the Twenty-First International Conference on
  Architectural Support for Programming Languages and Operating Systems},
  ASPLOS ’16, page 67–81, New York, NY, USA, 2016. Association for
  Computing Machinery.

\bibitem{pacingtoidle}
David H.~K. Kim, Connor Imes, and Henry Hoffmann.
\newblock Racing and pacing to idle: Theoretical and empirical analysis of
  energy optimization heuristics.
\newblock In {\em Proceedings of the 2015 IEEE 3rd International Conference on
  Cyber-Physical Systems, Networks, and Applications}, CPSNA '15, page 78–85,
  USA, 2015. IEEE Computer Society.

\bibitem{heuristics_0}
David~H.K. Kim, Connor Imes, and Henry Hoffmann.
\newblock Racing and pacing to idle: Theoretical and empirical analysis of
  energy optimization heuristics.
\newblock In {\em 2015 IEEE 3rd International Conference on Cyber-Physical
  Systems, Networks, and Applications}, pages 78--85, 2015.

\bibitem{4658633}
Wonyoung Kim, Meeta~S. Gupta, Gu-Yeon Wei, and David Brooks.
\newblock System level analysis of fast, per-core dvfs using on-chip switching
  regulators.
\newblock In {\em 2008 IEEE 14th International Symposium on High Performance
  Computer Architecture}, pages 123--134, 2008.

\bibitem{10.1145/1241601.1241609}
Masaaki Kondo, Hiroshi Sasaki, and Hiroshi Nakamura.
\newblock Improving fairness, throughput and energy-efficiency on a chip
  multiprocessor through dvfs.
\newblock {\em SIGARCH Comput. Archit. News}, 35(1):31–38, March 2007.

\bibitem{napsac}
Andrew Krioukov, Prashanth Mohan, Sara Alspaugh, Laura Keys, David Culler, and
  Randy Katz.
\newblock Napsac: Design and implementation of a power-proportional web
  cluster.
\newblock {\em SIGCOMM Comput. Commun. Rev.}, 41(1):102–108, January 2011.

\bibitem{slowdownorsleep}
Etienne Le~Sueur and Gernot Heiser.
\newblock Slow down or sleep, that is the question.
\newblock In {\em Proceedings of the 2011 USENIX Conference on USENIX Annual
  Technical Conference}, USENIXATC'11, page~16, USA, 2011. USENIX Association.

\bibitem{10.1145/1839667.1839670}
Benjamin~C. Lee and David Brooks.
\newblock Applied inference: Case studies in microarchitectural design.
\newblock {\em ACM Trans. Archit. Code Optim.}, 7(2), oct 2010.

\bibitem{10.1145/1168857.1168881}
Benjamin~C. Lee and David~M. Brooks.
\newblock Accurate and efficient regression modeling for microarchitectural
  performance and power prediction.
\newblock In {\em Proceedings of the 12th International Conference on
  Architectural Support for Programming Languages and Operating Systems},
  ASPLOS XII, page 185–194, New York, NY, USA, 2006. Association for
  Computing Machinery.

\bibitem{10.1145/1229428.1229479}
Benjamin~C. Lee, David~M. Brooks, Bronis~R. de~Supinski, Martin Schulz, Karan
  Singh, and Sally~A. McKee.
\newblock Methods of inference and learning for performance modeling of
  parallel applications.
\newblock In {\em Proceedings of the 12th ACM SIGPLAN Symposium on Principles
  and Practice of Parallel Programming}, PPoPP '07, page 249–258, New York,
  NY, USA, 2007. Association for Computing Machinery.

\bibitem{10.1145/1629911.1629926}
Jungseob Lee and Nam~Sung Kim.
\newblock Optimizing throughput of power- and thermal-constrained multicore
  processors using dvfs and per-core power-gating.
\newblock In {\em Proceedings of the 46th Annual Design Automation Conference},
  DAC '09, page 47–50, New York, NY, USA, 2009. Association for Computing
  Machinery.

\bibitem{4273098}
Charles Lefurgy, Xiaorui Wang, and Malcolm Ware.
\newblock Server-level power control.
\newblock In {\em Fourth International Conference on Autonomic Computing
  (ICAC'07)}, pages 4--4, 2007.

\bibitem{eurosys14}
Jacob Leverich and Christos Kozyrakis.
\newblock Reconciling high server utilization and sub-millisecond
  quality-of-service.
\newblock In {\em Proceedings of the Ninth European Conference on Computer
  Systems}, EuroSys '14, New York, NY, USA, 2014. Association for Computing
  Machinery.

\bibitem{10.1145/3342195.3387520}
Chi Li, Shu Wang, Henry Hoffmann, and Shan Lu.
\newblock Statically inferring performance properties of software
  configurations.
\newblock In {\em Proceedings of the Fifteenth European Conference on Computer
  Systems}, EuroSys '20, New York, NY, USA, 2020. Association for Computing
  Machinery.

\bibitem{1598114}
J.~Li and J.F. Martinez.
\newblock Dynamic power-performance adaptation of parallel computation on chip
  multiprocessors.
\newblock In {\em The Twelfth International Symposium on High-Performance
  Computer Architecture, 2006.}, pages 77--87, 2006.

\bibitem{mica}
Hyeontaek Lim, Dongsu Han, David~G. Andersen, and Michael Kaminsky.
\newblock {MICA}: A holistic approach to fast in-memory key-value storage.
\newblock In {\em 11th {USENIX} Symposium on Networked Systems Design and
  Implementation ({NSDI} 14)}, pages 429--444, Seattle, WA, April 2014.
  {USENIX} Association.

\bibitem{oldi-pegasus}
David Lo, Liqun Cheng, Rama Govindaraju, Luiz~Andr\'{e} Barroso, and Christos
  Kozyrakis.
\newblock Towards energy proportionality for large-scale latency-critical
  workloads.
\newblock In {\em Proceeding of the 41st Annual International Symposium on
  Computer Architecuture}, ISCA '14, page 301–312. IEEE Press, 2014.

\bibitem{heracles}
David Lo, Liqun Cheng, Rama Govindaraju, Parthasarathy Ranganathan, and
  Christos Kozyrakis.
\newblock Heracles: Improving resource efficiency at scale.
\newblock {\em SIGARCH Comput. Archit. News}, 43(3S):450–462, June 2015.

\bibitem{unikernels}
Anil Madhavapeddy, Richard Mortier, Charalampos Rotsos, David Scott, Balraj
  Singh, Thomas Gazagnaire, Steven Smith, Steven Hand, and Jon Crowcroft.
\newblock Unikernels: Library operating systems for the cloud.
\newblock In {\em Proceedings of the Eighteenth International Conference on
  Architectural Support for Programming Languages and Operating Systems},
  ASPLOS '13, pages 461--472, New York, NY, USA, 2013. ACM.

\bibitem{sandstorm}
Ilias Marinos, Robert~N.M. Watson, and Mark Handley.
\newblock Network stack specialization for performance.
\newblock In {\em Proceedings of the 2014 ACM Conference on SIGCOMM}, SIGCOMM
  '14, pages 175--186, New York, NY, USA, 2014. ACM.

\bibitem{powernap}
David Meisner, Brian~T. Gold, and Thomas~F. Wenisch.
\newblock Powernap: Eliminating server idle power.
\newblock In {\em Proceedings of the 14th International Conference on
  Architectural Support for Programming Languages and Operating Systems},
  ASPLOS XIV, page 205–216, New York, NY, USA, 2009. Association for
  Computing Machinery.

\bibitem{oldi-study}
David Meisner, Christopher~M. Sadler, Luiz~Andr\'{e} Barroso, Wolf-Dietrich
  Weber, and Thomas~F. Wenisch.
\newblock Power management of online data-intensive services.
\newblock In {\em Proceedings of the 38th Annual International Symposium on
  Computer Architecture}, ISCA '11, page 319–330, New York, NY, USA, 2011.
  Association for Computing Machinery.

\bibitem{dreamweaver}
David Meisner and Thomas~F. Wenisch.
\newblock Dreamweaver: Architectural support for deep sleep.
\newblock In {\em Proceedings of the Seventeenth International Conference on
  Architectural Support for Programming Languages and Operating Systems},
  ASPLOS XVII, page 313–324, New York, NY, USA, 2012. Association for
  Computing Machinery.

\bibitem{mellanoxsinterrupt}
{Mellanox}.
\newblock
  \url{https://community.mellanox.com/s/article/understanding-interrupt-moderation}.

\bibitem{caloree}
Nikita Mishra, Connor Imes, John~D. Lafferty, and Henry Hoffmann.
\newblock Caloree: Learning control for predictable latency and low energy.
\newblock {\em SIGPLAN Not.}, 53(2):184–198, mar 2018.

\bibitem{10.1145/2775054.2694373}
Nikita Mishra, Huazhe Zhang, John~D. Lafferty, and Henry Hoffmann.
\newblock A probabilistic graphical model-based approach for minimizing energy
  under performance constraints.
\newblock {\em SIGPLAN Not.}, 50(4):267–281, mar 2015.

\bibitem{amdenergy}
{Naveen Krishna Chatradhi }.
\newblock {Kernel driver amd\_energy}.
\newblock \url{https://www.kernel.org/doc/html/v5.9/hwmon/amd_energy.html}.

\bibitem{nature1}
{Nicola Jones}.
\newblock {How to stop data centres from gobbling up the world’s
  electricity}.
\newblock \url{https://www.nature.com/articles/d41586-018-06610-y}.

\bibitem{carat}
Adam~J. Oliner, Anand~P. Iyer, Ion Stoica, Eemil Lagerspetz, and Sasu Tarkoma.
\newblock Carat: Collaborative energy diagnosis for mobile devices.
\newblock In {\em Proceedings of the 11th ACM Conference on Embedded Networked
  Sensor Systems}, SenSys '13, New York, NY, USA, 2013. Association for
  Computing Machinery.

\bibitem{shenango}
Amy Ousterhout, Joshua Fried, Jonathan Behrens, Adam Belay, and Hari
  Balakrishnan.
\newblock Shenango: Achieving high cpu efficiency for latency-sensitive
  datacenter workloads.
\newblock In {\em Proceedings of the 16th USENIX Conference on Networked
  Systems Design and Implementation}, NSDI'19, page 361–377, USA, 2019.
  USENIX Association.

\bibitem{amdepp}
{Perry Yuan}.
\newblock {Implement AMD Pstate EPP Driver}.
\newblock \url{https://lwn.net/Articles/914431/}.

\bibitem{affinityaccept}
Aleksey Pesterev, Jacob Strauss, Nickolai Zeldovich, and Robert~T. Morris.
\newblock Improving network connection locality on multicore systems.
\newblock In {\em Proceedings of the 7th ACM European Conference on Computer
  Systems}, EuroSys ’12, page 337–350, New York, NY, USA, 2012. Association
  for Computing Machinery.

\bibitem{arrakis}
Simon Peter, Jialin Li, Irene Zhang, Dan R.~K. Ports, Doug Woos, Arvind
  Krishnamurthy, Thomas Anderson, and Timothy Roscoe.
\newblock Arrakis: The operating system is the control plane.
\newblock {\em ACM Trans. Comput. Syst.}, 33(4), November 2015.

\bibitem{powercap}
P.~{Petoumenos}, L.~{Mukhanov}, Z.~{Wang}, H.~{Leather}, and D.~S.
  {Nikolopoulos}.
\newblock Power capping: What works, what does not.
\newblock In {\em 2015 IEEE 21st International Conference on Parallel and
  Distributed Systems (ICPADS)}, pages 525--534, 2015.

\bibitem{flicker}
Paula Petrica, Adam~M. Izraelevitz, David~H. Albonesi, and Christine~A.
  Shoemaker.
\newblock Flicker: A dynamically adaptive architecture for power limited
  multicore systems.
\newblock {\em SIGARCH Comput. Archit. News}, 41(3):13–23, jun 2013.

\bibitem{mcdsilo}
George Prekas.
\newblock \url{https://github.com/ix-project/servers/tree/master}, 2017.

\bibitem{zygos}
George Prekas, Marios Kogias, and Edouard Bugnion.
\newblock Zygos: Achieving low tail latency for microsecond-scale networked
  tasks.
\newblock In {\em Proceedings of the 26th Symposium on Operating Systems
  Principles}, SOSP ’17, page 325–341, New York, NY, USA, 2017. Association
  for Computing Machinery.

\bibitem{ixcp}
George Prekas, Mia Primorac, Adam Belay, Christos Kozyrakis, and Edouard
  Bugnion.
\newblock Energy proportionality and workload consolidation for
  latency-critical applications.
\newblock In {\em Proceedings of the Sixth ACM Symposium on Cloud Computing},
  SoCC '15, page 342–355, New York, NY, USA, 2015. Association for Computing
  Machinery.

\bibitem{pytorchadam}
pytorch.
\newblock {Adam}.
\newblock
  \url{https://pytorch.org/docs/stable/generated/torch.optim.Adam.html}.

\bibitem{arachne}
Henry Qin, Qian Li, Jacqueline Speiser, Peter Kraft, and John Ousterhout.
\newblock Arachne: Core-aware thread management.
\newblock In {\em 13th {USENIX} Symposium on Operating Systems Design and
  Implementation ({OSDI} 18)}, pages 145--160, Carlsbad, CA, October 2018.
  {USENIX} Association.

\bibitem{cpuidle_policy}
{Rafael J. Wysocki}.
\newblock {CPU Idle Time Management}.
\newblock
  \url{https://www.kernel.org/doc/html/v5.0/admin-guide/pm/cpuidle.html}.

\bibitem{scalingmcdfacebook}
{Rajesh Nishtala and Hans Fugal and Steven Grimm and Marc Kwiatkowski and
  Herman Lee and Harry C. Li and Ryan McElroy and Mike Paleczny and Daniel Peek
  and Paul Saab and David Stafford and Tony Tung and Venkateshwaran
  Venkataramani}.
\newblock {Scaling Memcache at Facebook}.
\newblock In {\em Presented as part of the 10th {USENIX} Symposium on Networked
  Systems Design and Implementation ({NSDI} 13)}, pages 385--398, Lombard, IL,
  2013. {USENIX}.

\bibitem{aliraza}
Ali Raza, Parul Sohal, James Cadden, Jonathan Appavoo, Ulrich Drepper, Richard
  Jones, Orran Krieger, Renato Mancuso, and Larry Woodman.
\newblock Unikernels: The next stage of linux’s dominance.
\newblock In {\em Proceedings of the Workshop on Hot Topics in Operating
  Systems}, HotOS ’19, page 7–13, New York, NY, USA, 2019. Association for
  Computing Machinery.

\bibitem{linux-core-ops}
Xiang~(Jenny) Ren, Kirk Rodrigues, Luyuan Chen, Camilo Vega, Michael Stumm, and
  Ding Yuan.
\newblock An analysis of performance evolution of linux's core operations.
\newblock In {\em Proceedings of the 27th ACM Symposium on Operating Systems
  Principles}, SOSP '19, page 554–569, New York, NY, USA, 2019. Association
  for Computing Machinery.

\bibitem{10.1145/3453483.3454109}
Rohan~Basu Roy, Tirthak Patel, Vijay Gadepally, and Devesh Tiwari.
\newblock Bliss: Auto-tuning complex applications using a pool of diverse
  lightweight learning models.
\newblock In {\em Proceedings of the 42nd ACM SIGPLAN International Conference
  on Programming Language Design and Implementation}, PLDI 2021, page
  1280–1295, New York, NY, USA, 2021. Association for Computing Machinery.

\bibitem{10.5555/2523721.2523732}
Hiroshi Sasaki, Satoshi Imamura, and Koji Inoue.
\newblock Coordinated power-performance optimization in manycores.
\newblock In {\em Proceedings of the 22nd International Conference on Parallel
  Architectures and Compilation Techniques}, PACT '13, page 51–62. IEEE
  Press, 2013.

\bibitem{ebbrt}
Dan Schatzberg, James Cadden, Han Dong, Orran Krieger, and Jonathan Appavoo.
\newblock Ebbrt: A framework for building per-application library operating
  systems.
\newblock In {\em 12th USENIX Symposium on Operating Systems Design and
  Implementation (OSDI 16)}, pages 671--688, GA, 2016. USENIX Association.

\bibitem{271072}
Michael~W. Shaffer.
\newblock A linux appliance construction set.
\newblock In {\em 14th Systems Administration Conference (LISA 2000)}, New
  Orleans, LA, December 2000. USENIX Association.

\bibitem{snell1996netpipe}
Quinn~O Snell, Armin~R Mikler, and John~L Gustafson.
\newblock {Netpipe: A Network Protocol Independent Performance Evaluator}.
\newblock In {\em IASTED International Conference on Intelligent Information
  Management and Systems}, 1996.

\bibitem{koala}
David~C. Snowdon, Etienne Le~Sueur, Stefan~M. Petters, and Gernot Heiser.
\newblock Koala: A platform for os-level power management.
\newblock In {\em Proceedings of the 4th ACM European Conference on Computer
  Systems}, EuroSys '09, page 289–302, New York, NY, USA, 2009. Association
  for Computing Machinery.

\bibitem{10.1109/IGCC.2011.6008552}
V.~Spiliopoulos, S.~Kaxiras, and G.~Keramidas.
\newblock Green governors: A framework for continuously adaptive dvfs.
\newblock In {\em Proceedings of the 2011 International Green Computing
  Conference and Workshops}, IGCC '11, page 1–8, USA, 2011. IEEE Computer
  Society.

\bibitem{NLP-energy}
Emma Strubell, Ananya Ganesh, and Andrew McCallum.
\newblock Energy and policy considerations for deep learning in {NLP}.
\newblock In {\em Proceedings of the 57th Annual Meeting of the Association for
  Computational Linguistics}, pages 3645--3650, Florence, Italy, July 2019.
  Association for Computational Linguistics.

\bibitem{twine}
Chunqiang Tang, Kenny Yu, Kaushik Veeraraghavan, Jonathan Kaldor, Scott
  Michelson, Thawan Kooburat, Aravind Anbudurai, Matthew Clark, Kabir Gogia,
  Long Cheng, Ben Christensen, Alex Gartrell, Maxim Khutornenko, Sachin
  Kulkarni, Marcin Pawlowski, Tuomas Pelkonen, Andre Rodrigues, Rounak
  Tibrewal, Vaishnavi Venkatesan, and Peter Zhang.
\newblock Twine: A unified cluster management system for shared infrastructure.
\newblock In {\em 14th {USENIX} Symposium on Operating Systems Design and
  Implementation ({OSDI} 20)}, pages 787--803. {USENIX} Association, November
  2020.

\bibitem{4061117}
Gerald Tesauro.
\newblock Reinforcement learning in autonomic computing: A manifesto and case
  studies.
\newblock {\em IEEE Internet Computing}, 11(1):22--30, 2007.

\bibitem{hotpower2008}
Niraj Tolia, Zhikui Wang, Manish Marwah, Cullen Bash, Parthasarathy
  Ranganathan, and Xiaoyun Zhu.
\newblock Delivering energy proportionality with non energy-proportional
  systems: Optimizing the ensemble.
\newblock In {\em Proceedings of the 2008 Conference on Power Aware Computing
  and Systems}, HotPower'08, page~2, USA, 2008. USENIX Association.

\bibitem{268014}
Niraj Tolia, Zhikui Wang, Manish Marwah, Cullen Bash, Parthasarathy
  Ranganathan, and Xiaoyun Zhu.
\newblock Delivering energy proportionality with non energy-proportional
  systems{\textemdash}optimizing the ensemble.
\newblock In {\em Workshop on Power Aware Computing and Systems (HotPower 08)},
  San Diego, CA, December 2008. {USENIX} Association.

\bibitem{10.1145/2829950}
Erik Tomusk, Christophe Dubach, and Michael O’boyle.
\newblock Four metrics to evaluate heterogeneous multicores.
\newblock {\em ACM Trans. Archit. Code Optim.}, 12(4), nov 2015.

\bibitem{silo}
Stephen Tu, Wenting Zheng, Eddie Kohler, Barbara Liskov, and Samuel Madden.
\newblock Speedy transactions in multicore in-memory databases.
\newblock In {\em Proceedings of the Twenty-Fourth ACM Symposium on Operating
  Systems Principles}, SOSP ’13, page 18–32, New York, NY, USA, 2013.
  Association for Computing Machinery.

\bibitem{10.1145/2830772.2830779}
Balajee Vamanan, Hamza~Bin Sohail, Jahangir Hasan, and T.~N. Vijaykumar.
\newblock Timetrader: Exploiting latency tail to save datacenter energy for
  online search.
\newblock In {\em Proceedings of the 48th International Symposium on
  Microarchitecture}, MICRO-48, page 585–597, New York, NY, USA, 2015.
  Association for Computing Machinery.

\bibitem{10.1145/3035918.3064029}
Dana Van~Aken, Andrew Pavlo, Geoffrey~J. Gordon, and Bohan Zhang.
\newblock Automatic database management system tuning through large-scale
  machine learning.
\newblock In {\em Proceedings of the 2017 ACM International Conference on
  Management of Data}, SIGMOD '17, page 1009–1024, New York, NY, USA, 2017.
  Association for Computing Machinery.

\bibitem{10.1145/3173162.3173206}
Shu Wang, Chi Li, Henry Hoffmann, Shan Lu, William Sentosa, and Achmad~Imam
  Kistijantoro.
\newblock {\em Understanding and Auto-Adjusting Performance-Sensitive
  Configurations}, page 154–168.
\newblock Association for Computing Machinery, New York, NY, USA, 2018.

\bibitem{seda}
Matt Welsh, David Culler, and Eric Brewer.
\newblock Seda: An architecture for well-conditioned, scalable internet
  services.
\newblock {\em SIGOPS Oper. Syst. Rev.}, 35(5):230–243, October 2001.

\bibitem{7851506}
Jonathan~A. Winter, David~H. Albonesi, and Christine~A. Shoemaker.
\newblock Scalable thread scheduling and global power management for
  heterogeneous many-core architectures.
\newblock In {\em 2010 19th International Conference on Parallel Architectures
  and Compilation Techniques (PACT)}, pages 29--39, 2010.

\bibitem{Wu2023}
Nan Wu and Yuan Xie.
\newblock A survey of machine learning for computer architecture and systems.
\newblock {\em {ACM} Computing Surveys}, 55(3):1--39, apr 2023.

\bibitem{Dynamo}
Qiang Wu, Qingyuan Deng, Lakshmi Ganesh, Chang-Hong Hsu, Yun Jin, Sanjeev
  Kumar, Bin Li, Justin Meza, and Yee~Jiun Song.
\newblock Dynamo: Facebook's data center-wide power management system.
\newblock In {\em Proceedings of the 43rd International Symposium on Computer
  Architecture}, ISCA '16, pages 469--480, Piscataway, NJ, USA, 2016. IEEE
  Press.

\bibitem{6493638}
Weidan Wu and Benjamin~C. Lee.
\newblock Inferred models for dynamic and sparse hardware-software spaces.
\newblock In {\em 2012 45th Annual IEEE/ACM International Symposium on
  Microarchitecture}, pages 413--424, 2012.

\bibitem{twittermcd}
Juncheng Yang, Yao Yue, and K.~V. Rashmi.
\newblock A large scale analysis of hundreds of in-memory cache clusters at
  twitter.
\newblock In {\em 14th USENIX Symposium on Operating Systems Design and
  Implementation (OSDI 20)}, pages 191--208. USENIX Association, November 2020.

\bibitem{6730744}
Nezih Yigitbasi, Theodore~L. Willke, Guangdeng Liao, and Dick Epema.
\newblock Towards machine learning-based auto-tuning of mapreduce.
\newblock In {\em 2013 IEEE 21st International Symposium on Modelling, Analysis
  and Simulation of Computer and Telecommunication Systems}, pages 11--20,
  2013.

\bibitem{7425206}
Xin Zhan, Reza Azimi, Svilen Kanev, David Brooks, and Sherief Reda.
\newblock Carb: A c-state power management arbiter for latency-critical
  workloads.
\newblock {\em IEEE Computer Architecture Letters}, 16(1):6--9, 2017.

\bibitem{10.1145/2872362.2872375}
Huazhe Zhang and Henry Hoffmann.
\newblock Maximizing performance under a power cap: A comparison of hardware,
  software, and hybrid techniques.
\newblock In {\em Proceedings of the Twenty-First International Conference on
  Architectural Support for Programming Languages and Operating Systems},
  ASPLOS '16, page 545–559, New York, NY, USA, 2016. Association for
  Computing Machinery.

\bibitem{7551432}
Yanqi Zhou, Henry Hoffmann, and David Wentzlaff.
\newblock Cash: Supporting iaas customers with a sub-core configurable
  architecture.
\newblock In {\em 2016 ACM/IEEE 43rd Annual International Symposium on Computer
  Architecture (ISCA)}, pages 682--694, 2016.

\bibitem{6522303}
Yuhao Zhu and Vijay~Janapa Reddi.
\newblock High-performance and energy-efficient mobile web browsing on
  big/little systems.
\newblock In {\em 2013 IEEE 19th International Symposium on High Performance
  Computer Architecture (HPCA)}, pages 13--24, 2013.

\bibitem{bestconfig}
Yuqing Zhu, Jianxun Liu, Mengying Guo, Yungang Bao, Wenlong Ma, Zhuoyue Liu,
  Kunpeng Song, and Yingchun Yang.
\newblock Bestconfig: Tapping the performance potential of systems via
  automatic configuration tuning.
\newblock In {\em Proceedings of the 2017 Symposium on Cloud Computing}, SoCC
  '17, page 338–350, New York, NY, USA, 2017. Association for Computing
  Machinery.

\end{thebibliography}
